%% file: main.tex
\documentclass[acmsmall]{acmart}
  
\usepackage{lineno,hyperref}

\usepackage{algorithm} 
\usepackage{algpseudocode}
\usepackage{graphicx}
\usepackage{textcomp}
\usepackage{xcolor}
\usepackage{comment}
\usepackage{breakurl}
\usepackage{xspace}
\usepackage{subfigure}
\usepackage{pifont}
\usepackage{multirow}
\usepackage{amsfonts}
\usepackage{listings}
\usepackage{subcaption}
\usepackage{color}
\usepackage{url}
\usepackage{booktabs}
\usepackage{colortbl}
\usepackage{tcolorbox}
\usepackage{hyperref}
\usepackage{ulem}
\usepackage{fvextra}
\usepackage{enumitem}
\usepackage{listings}

\usepackage{cases}
\usepackage{textcomp}

\definecolor{light-gray}{gray}{0.95}

\newcommand{\tool}{\textsf{ShortenDoc}}
\newcommand{\code}[1]{\colorbox{light-gray}{\texttt{#1}}}

\definecolor{dkgreen}{rgb}{0,0.6,0}
\definecolor{gray}{rgb}{0.5,0.5,0.5}
\definecolor{mauve}{rgb}{0.58,0,0.82}

\newcommand{\yg}[1]{\textcolor{black}{#1}}

\AtBeginDocument{%
  \providecommand\BibTeX{{%
    \normalfont B\kern-0.5em{\scshape i\kern-0.25em b}\kern-0.8em\TeX}}}

\setcopyright{acmcopyright}
\copyrightyear{}
\acmYear{}
\acmDOI{}

\acmVolume{0}
\acmNumber{0}
\acmArticle{0}
\acmMonth{0}

\begin{document}

\title{Less is More: DocString Compression in Code Generation}

\author{Guang Yang}
\email{novelyg@outlook.com}
\author{Yu Zhou}
\authornote{Corresponding author.}
\email{zhouyu@nuaa.edu.cn}
\author{Wei Cheng}
\email{chengweii@nuaa.edu.cn}
\author{Xiangyu Zhang}
\email{zhangx1angyu@nuaa.edu.cn}
\affiliation{
\institution{Nanjing University of Aeronautics and Astronautics}
\city{Nanjing}
\country{China}
}
\author{Xiang Chen}
\email{xchencs@ntu.edu.cn}
\affiliation{
\institution{Nantong University}
\city{Nantong}
\country{China}
}
\author{Terry Yue Zhuo}
\email{terry.zhuo@monash.edu}
\affiliation{%
	\institution{Monash University}
	\country{Australia}
}
\author{Ke Liu}
\email{liuke23@nudt.edu.cn}
\affiliation{%
\institution{National University of Defense Technology}
\city{ChangSha}
\country{China}
}
\author{Xin Zhou}
\email{xinzhou.2020@phdcs.smu.edu.sg}
\author{David Lo}
\email{davidlo@smu.edu.sg}
\affiliation{
\institution{Singapore Management University}
\country{Singapore}
}
\author{Taolue Chen}
\email{t.chen@bbk.ac.uk}
\authornote{Corresponding author.}
\affiliation{
\institution{Birkbeck, University of London}
\city{London}
\country{UK}
}

\renewcommand{\shortauthors}{G. Yang, et al.}

%%%%%%%%%%%%%%%%%%%%%%%%%%%%%%%%%%%%%%%%%%%%%%%%%%%%%%%%%%%%%%%%%%%%%%%%%%%%%%%

\begin{abstract}
The widespread use of Large Language Models (LLMs) in software engineering has intensified the need for improved model and resource efficiency. In particular, for neural code generation, LLMs are used to translate function/method signature and DocString to executable code.
DocStrings, which capture user requirements for the code and are typically used as the prompt for LLMs, often contain redundant information. Recent advancements in prompt compression have shown promising results in Natural Language Processing (NLP), but their applicability to code generation remains uncertain. Our empirical study show that the state-of-the-art prompt compression methods achieve only about 10\% reduction, as further reductions would cause significant performance degradation. In our study, we propose a novel compression method, {\tool}, dedicated to DocString compression for code generation. Our experiments on six code generation datasets, 
five open-source LLMs (1B to 10B parameters) and one closed-source LLM GPT-4o confirm that {\tool} achieves 25--40\% compression while preserving the quality of generated code, outperforming other baseline methods at similar compression levels. The benefit of this method is to improve efficiency and reduce the token processing cost while maintaining the quality of the generated code, especially when calling third-party APIs. %, and is able to reduce the token processing cost by 25--40\%.
\end{abstract}
%%%%%%%%%%%%%%%%%%%%%%%%%%%%%%%%%%%%%%%%%%%%%%%%%%%%%%%%%%%%%%%%%%%%%%%%%%%%%%%




\maketitle
% \linenumbers
\input{sections/1intro}
\input{sections/2background}
\input{sections/3empirical}
\input{sections/4approach}

\input{sections/5results}
\input{sections/6discussion}
\input{sections/7relate}

\input{sections/8conclusion}

\begin{acks}
This work was partially supported by the National Natural Science Foundation of China (NSFC, No.\ 62372232), the Collaborative Innovation Center of Novel Software Technology and Industrialization, and the Short-term Visiting Program of Nanjing University of Aeronautics and Astronautics for Ph.D. Students Abroad (No.\ 240602DF16), and the A*STAR under its 2nd CSIRO and A*STAR: Research-Industry (2+2) Partnership Program (Award R24I5IR047). Any opinions, findings and conclusions or recommendations expressed in this material are those of the author(s) and do not reflect the views of the A*STAR.
T.\ Chen is partially supported by an oversea grant from the State Key Laboratory of Novel Software Technology, Nanjing University (KFKT2023A04).
\end{acks}
 
\bibliographystyle{Reference}
\bibliography{acmart}
 
\end{document}

%% file: sections/1intro.tex
\section{Introduction}

% 引言
% 1、引入大型语言模型（LLMs）在软件工程中的重要性。
% 2、强调代码生成任务中模型效率和成本优化的需求。
Large Language Models (LLMs) have emerged as a key tool in today's software development, %, for driving technological innovation and productivity~\cite{nam2024using, rasnayaka2024empirical}. 
%These models, through 
owning to their powerful language understanding and generation capabilities~\cite{nam2024using, rasnayaka2024empirical}. In particular, they have shown great potential in a variety of SE tasks such as code generation~\cite{liu2024your}, automated testing~\cite{chen2024chatunitest} and documentation~\cite{dvivedi2024comparative}.
% 
%The ability of LLMs to understand and generate syntactically and logically compliant code greatly simplifies the software development process and improves development efficiency.
However, with the wide deployment of LLMs, it has become a pressing challenge to improve the inference efficiency and to reduce the ever increasing demand of computational resources~\cite{niu2024evaluating,shi2024greening,yang2024robustness}. 
This challenge manifests in two aspects: (1) the computational overhead in terms of FLOPs during model inference, and (2) the financial cost associated with API calls to third-party LLMs for code generation.

%the efficiency~\cite{niu2024evaluating} and cost optimization~\cite{shi2024greening} of the models have become a pressing issue. 
%
%This not only puts higher demands on the computational resources of the model, but also increases the inference cost~\cite{yang2024robustness}. 
%Therefore, how to optimize the efficiency of the model and reduce the cost while maintaining the quality of the generated code is an important research topic in the field of software engineering.
% \begin{figure}[ht]
% \centering
% \includegraphics[width=0.9\textwidth]{figs/DocString.png}
% \caption{The Example of DocString in Code Generation \label{fig:docstring}}
% \vspace{-0.3cm}
% \end{figure}

\begin{lstlisting}[
caption={The Example of DocString in Code Generation}, 
label=lst:docstring
]
def add_numbers(a, b):
    """
    Adds two numbers and returns the result.
        Parameters:
    a (int or float): The first number to add.
    b (int or float): The second number to add.
        Returns:
    int or float: The sum of a and b.
        Raises:
    TypeError: If 'a' or 'b' is not a number.
        Example:
    >>> add_numbers(2, 3)
    5
    """
    if not (isinstance(a, (int, float)) and isinstance(b, (int, float))):
        raise TypeError("Both arguments must be numbers")
    return a + b
\end{lstlisting}

% 背景
% 1、描述DocString在代码生成中的作用和重要性。
% 2、讨论现有DocString的质量差异及其对模型性能和成本的影响。
DocString (documentation string), a string literal embedded in source code to document specific features or functionalities, serves as a critical component in software development. It significantly enhances code readability, maintainability and usability. Typically positioned at the beginning of a function or method, DocString plays a pivotal role in the code generation process by articulating the code's functionality, parameters, return values and potential exceptions that may arise~\cite{dainese2024can}.
Listing~\ref{lst:docstring} presents an example to illustrate how DocString (Lines 2-14) elucidates the purpose of the function, its parameters, the type of return value it yields, the exceptions it might trigger, and offers a practical example showing how to invoke the function. 
% These texts not only facilitate an intuitive grasp of the code for developers but also act as a direct reflection of user requirements.

\smallskip
\noindent\textbf{Motivation.}
In the realm of code generation leveraging LLMs, DocStrings are typically used as prompts. They guide the model to generate code that aligns with the document specific features or functionalities~\cite{miceli2017parallel}. 
However, existing DocString varies significantly in quality~\cite{poudel2024documint}. 
Some DocString may be overly lengthy and contain a large amount of redundancy, which may prevent the model from accurately understanding user requirements. 
Although these redundant DocString do not exceed the length of the maximum context that LLM can now handle, they may also increase the computational cost of the model inference,  thus reduce their efficiency. 
For instance, the computational cost of the model inference is usually proportional to the length of the input prompt. The longer the DocString, the higher the computational cost of model inference. 
According to OpenAI founder Sam Altman, GPT-4o processes approximately 200 billion tokens daily\footnote{\url{https://x.com/sama/status/1815437745550172617}}, while Baidu CEO Robin Li revealed (at Baidu World 2024) that ERNIE model handles over 1.5 billion calls per day\footnote{\url{https://ir.baidu.com/news-releases/news-release-details/baidu-announces-third-quarter-2024-results/}}. 
At such massive processing scales, even a modest 25\%-40\% compression of user inputs could significantly reduce overall computational costs and energy consumption in the long run.

In addition, when users generate code by invoking APIs from third-party LLMs, shorter prompt can greatly reduce their financial cost. 
Therefore, optimizing the quality of DocString is one of the key factors for improving the efficiency of code generation and reducing costs.
For instance, the costs of LLM API are associated with input tokens (i.e., prefilling) and output tokens (i.e., decoding), where input tokens are usually 1--5 times cheaper than output tokens (varying across different LLM providers). 
For example, Meta's Llama 3.1 405B model charges at the same rate for both input and output tokens\footnote{\url{https://llama3-1.com/price/}}; OpenAI's models price input tokens at 1/4 of output tokens\footnote{\url{https://openai.com/api/pricing/}}; Anthropic's models set input tokens at 1/5 of output token costs\footnote{\url{https://www.anthropic.com/pricing}}.
Assuming one usage consumes 512 input tokens and 512 output tokens on average, with a 50\% input token reduction, the end-to-end cost saving is $(512 \times 1 \times 50\%) / (512 \times 1 + 512 \times 4) = 10\%$ according to OpenAI's pricing model, and 25\% according to Meta's pricing model. 
While the cost reduction for a single API call might seem modest due to the pricing difference between input and output tokens across LLM providers, the cumulative economic benefits become substantial in large-scale deployment scenarios or when synthesizing large-scale code generation datasets using LLMs.

Moreover, DocString compression can be complementary to other efficiency-oriented techniques in code generation, such as efficient decoding strategies~\cite{sun2024neural, sun2024don} and optimized programming language grammar~\cite{sun2024ai}. 
Meanwhile, the insights gained from DocString compression in zero-shot code generation scenarios can also provide theoretical foundations for more complex retrieval-augmented generation (RAG) based code generation tasks.

% 文献回顾 & 问题陈述
% 1、概述自然语言处理（NLP）中提示压缩的进展。
% 2、分析这些压缩技术在代码生成中的局限性。

Naturally, one may apply existing prompt compression techniques (e.g., Selective\_Context~\cite{li2023compressing} and LLMLingua~\cite{jiang2023llmlingua}) from NLP to DocString.
%These techniques have been successful in improving model inference speed by reducing input length, lowering costs, and achieving certain successes in specific NLP tasks.
%However, when attempting to apply these techniques to code generation tasks, we encountered 
It turns out that there are several challenges: 

\textbf{(1) Compression efficacy}: 
these compression methods show limited effectiveness in compressing DocStrings. 
%They struggle to find a balance between the compression rate and the quality of the generated code.
As evidenced by our experimental results in Section~\ref{sec:empirical} (Table~\ref{tab:em}), increasing the compression rate beyond 10\% may incur a significant decrease in the quality of the generated code.
% When we attempted to increase the compression rate above 10\%, there was a significant decrease in the quality of the generated code. %, indicating that existing techniques are inadequate in preserving critical information.
%
We posit that this is largely because %largely due to the fact that current compression techniques often 
they fail to %account for 
extract the code-related semantic information in DocStrings, %. Compression methods that lack an understanding of code semantics may 
leading to significant information loss. % of essential information, which in turn affects the quality and usability of the code.

\textbf{(2) Inflexibility}: 
existing methods require manual setting of the compression ratio, %which not only increases the complexity of the operation but also 
making it difficult to adapt to various code generation scenarios. 
Developers must adjust compression parameters based on specific situations, which requires specialized knowledge and is time-consuming. Moreover, it is hard to find the optimal balance of compression ratio and model efficacy (specifically Pass@1, which measures the percentage of tasks for which the model produces a correct output on its first attempt). 
%Moreover, the manually set compression ratio may not accurately align with the model's optimal working point, resulting in a compressed DocString that does not maximize the efficiency of code generation.

% 研究方法
\smallskip
\noindent\textbf{Our work.}
%In this research, our primary objective is 
We aim to develop prompt compression methods dedicated to DocStrings, %in code prompts while preserving their semantic integrity, thereby 
optimizing the efficiency of LLMs in code generation tasks. 
We introduce {\tool}, a novel method that dynamically adjusts DocString compression based on the importance of individual tokens. 
Instead of using a fixed compression ratio, {\tool} analyzes the significance of each token, ensuring that essential information is preserved while removing redundant content.

Initially, we decompose a given prompt into its signature and DocString components. 
The DocString undergoes preprocessing to enhance its quality based on empirical findings. 
Each token in the DocString is assigned an importance score, and tokens are ranked accordingly. 
We then construct a search space of tokens eligible for compression, considering predefined constraints to ensure minimal semantic distortion. 

Tokens are iteratively compressed until the constraint is no longer satisfied. 
The final compressed DocString is obtained by integrating the selected tokens with the original sequence.
This approach ensures that {\tool} balances efficiency with accuracy, minimizing computational overhead while maintaining the quality of the generated code across different programming languages and contexts.

% 主要结果
%Through comprehensive experiments, 
We evaluate the performance of {\tool} across six datasets and six LLMs, demonstrating its superior ability to preserve essential information and reduce inference costs compared to baseline methods. 
To explore the generalization of {\tool}, we evaluate the performance of {\tool} across four more programming languages.
Additionally, we conduct a human study to further assess the practical impact of DocString compression, evaluating both informativeness and comprehensibility from the perspective of human evaluation.
% Finally, we delve into the insights gained from DocString compression techniques and give relevant insights.

Our contributions can be summarized as follows. 

\begin{itemize}  
\item We empirically show the feasibility and limitations of DocString compression for code generation tasks.

\item We design a novel method {\tool}, which is an adaptive compression method for compressing DocString and has better compression results compared to existing methods.

\item We delve into the insights gained from DocString compression techniques and give relevant insights.

\end{itemize}

%%%%%%%%%%%%%%%%%%%%%%%%%%%%%%%%%%%%%%%%%%%%%%%%%%%%%%%%%%%%%%%%%%%%%%%%%%%%%

\noindent{\bf Structure.} The rest of the paper is organized as follows. 
Section~\ref{sec:back} provides preliminary knowledge related to our study.
Section~\ref{sec:empirical} confirms and analyzes the feasibility and limitations of existing DocString compression methods for code generation tasks.
Section~\ref{sec:method} describes the key components of {\tool}. 
Section~\ref{sec:evaluation} present the research questions and the result analysis, which is followed by the discussion in Section~\ref{sec:discuss}. % and potential threat to the validity of our empirical study. 
Section~\ref{sec:related} reviewes the related work.
Section~\ref{sec:conclusion} concludes our study and outlines future directions.

\yg{To facilitate reproducibility, all source code and experimental data are released at \url{https://github.com/NTDXYG/ShortenDoc}}.

%% file: sections/2background.tex
\section{Background}
\label{sec:back}

Code generation, in a nutshell, refers to automated generation of code from specifications and under certain constraints, which plays a pivotal role in software development~\cite{li2022competition}. 
In this context, signatures and DocStrings (in natural language) are typically used as inputs to the model and are converted into executable code.

Let $\mathcal{T}$ denote the set of signatures, $\mathcal{D}$ denote the set of DocStrings and $\mathcal{Y}$ denote the set of executable code. In general, signatures define the name of the function/method, input parameters, return type and possible side effects.
DocStrings provide a natural language description of the function or method, including its purpose, behavior, parameter explanations, and return values.
Typically, code generation can be formalized as a function $f: \mathcal{T}\times \mathcal{D} \rightarrow \mathcal{Y}$. 

% \tl{correct the math}
In practice, most of the current code generation models follow Transformer's decoder architecture~\cite{DBLP:conf/nips/VaswaniSPUJGKP17}. 
The process can be broken down into several steps, including word embedding, layer transformations and   vocabulary mapping. 

%Here, we provide a detailed formalization of these steps, culminating in the probability of the generated code given the input specifications.

%%%%%%%%%%%%%%%%%%%%%%%%%%%%%%%%%%%%%%%%%%%%%%%%%%%%%%%%%%

\noindent\textbf{Embedding.} The input signature $T\in \mathcal{T}$ and DocString $D\in \mathcal{D}$ are tokenized into a sequence of words \( w_1,\ldots, w_n \). 
Each word is then converted into a vector representation through word embedding.
Let $\mathcal{W}$ be the vocabulary of tokens and $\mathbf{E} \in \mathbb{R}^{|\mathcal{W}| \times d_{\text{embed}}}$ be the embedding matrix, where $d_{\text{embed}}$ is the dimensionality of the embedding space. 
The word embedding function $\text{Emb}: \mathcal{W} \rightarrow \mathbb{R}^{d_{\text{embed}}}$ maps each token to its corresponding vector representation, viz., \( \text{Emb}(w_i) \) is the embedding vector for the \( i \)-th word.

\noindent\textbf{Transformer.} The embedded sequence is then fed into a stack of \( L \) transformer layers to produce a sequence of hidden states \( \mathbf{H} = [\mathbf{h}_1, \mathbf{h}_2, \ldots, \mathbf{h}_L] \), where \( \mathbf{h}_i \) is the hidden state vector for the \( i \)-th layer output and each \( \text{Transformer-Block} \) computes the hidden state through Self-Attention and Feed-Forward, i.e., 
\[
\mathbf{h}_{i+1} = \text{Transformer-Block}(\mathbf{h}_i) = \text{Self-Attention}(\mathbf{h}_i) + \text{Feed-Forward}(\mathbf{h}_i)
\]
% where each \( \text{Transformer-Block} \) computes the hidden state as follows:
% \[
% \mathbf{h}_{i+1} = \text{Self-Attention}(\mathbf{h}_i) + \text{Feed-Forward}(\mathbf{h}_i)
% \]

\noindent\textbf{Probability.} 
The final hidden state vector \( \mathbf{h}_L \) of the last layer is mapped to a vocabulary space using a linear transformation, followed by a softmax function to obtain a probability distribution over the vocabulary: $\mathbf{z} = \mathbf{W} \mathbf{h}_L + \mathbf{b}$, where \( \mathbf{W} \) is the weight matrix and \( \mathbf{b} \) is the bias vector for the vocabulary mapping.

The probability of generating the next code token \( y \) is computed using the softmax function:
\[
P(y) = \text{softmax}(\mathbf{z}) = \frac{\exp(\mathbf{z}_y)}{\sum_{j=1}^{|\mathcal{W}|} \exp(\mathbf{z}_j)}
\]
where \( \mathbf{z}_y \) is the score for the word \( y \) in the vocabulary, and \( |\mathcal{W}| \) is the size of the vocabulary.
\[
P(Y|T, D) = \prod_{i=1}^{m}P(y_{i}|T, D, y_{1}, y_{2}, \cdots, y_ {i-1})
\]
The model generates a sequence of code tokens \( y_1, y_2, \ldots, y_m \) by sampling from the probability distributions \( P(y_1|T, D), P(y_2|T, D, y_1), \ldots, P(y_m|T, D, y_1, \ldots, y_{m-1}) \), where each token is conditioned on the previous tokens in the sequence.

%%%%%%%%%%%%%%%%%%%%%%%%%%%%%%%%%%%%%%%%%%%%%%%%%%%%%%%%%%
\noindent\textbf{Prompt Compression.}
Prompt compression is a technique aimed at reducing the length of input prompts while preserving essential information.  % to enhance the performance of the model. 
This method is particularly beneficial in scenarios where computational resources are constrained or where rapid response time is critical. 

%In this process, we can define the original prompt \( P \) and the compressed prompt \( P' \) as follows:
In a nutshell, the goal of prompt compression is to construct a function \( g \), which, given the original prompt $P$, outputs  the compressed prompt $g(P)$ such that % that ensures the compressed prompt \( P' \) 
the it contains less tokens but retains the critical information.  %from the original prompt \( P \) while reducing its length:
%\begin{equation}
%	P' = g(P)
%\end{equation}

%\[
%P = \{p_1, p_2, \ldots, p_k\}, \quad \text{where} \ k = |P|
%\]
%\[
%P' = \{p'_1, p'_2, \ldots, p'_{k'}\}, \quad \text{where} \ k' = |P'|~\text{and}~k' < k
%\]
%Here, the $\{p_1, p_2, \ldots, p_k\}$ represents the subtokens splited by a special pretrained tokenizer. 
%
%The mapping function $g$ is implemented according to different compression strategies.
%In our study, one of our main objectives is to research suitable function $g$ for code generation tasks.
%
%\tl{this formalization is trivial.}

%% file: sections/3empirical.tex
\section{Empirical Study}
\label{sec:empirical}

In this section, we conduct an empirical study to explore the applicability of existing prompt compression methods to code generation tasks.

\subsection{Experiment Setup}

\noindent{\bf Datasets.} 
% We experiment on six diverse and widely adopted datasets %that are designed to 
% which can simulate real-world code generation tasks. %These datasets vary in complexity and domain, providing a comprehensive evaluation. %evaluating the effectiveness of our compression techniques.
We experiment on six diverse and widely adopted datasets which can simulate real-world code generation tasks. 
Our dataset selection follows two principles: (1) including dataset variants derived from the same base to evaluate robustness, and (2) covering different sources and complexity levels to ensure diversity.

For the first principle, we select HumanEval and its two variants from EvoEval (Subtle and Creative). 
While these datasets share similar base problems, they differ in how requirements are presented. This setup allows us to evaluate {\tool}'s robustness in handling different expressions of the same functional requirement.

For the second principle, we select datasets from varied sources and at different complexity levels: MBPP contains human-written entry-level problems, CodeHarmony provides LLM-synthesized high-quality samples, and BigCodeBench presents practical and challenging programming tasks. The diversity ensures a comprehensive evaluation across different scenarios. In particular, 

\begin{itemize}
\item \textbf{HumanEval~\cite{chen2021evaluating}:} this dataset contains 164 well-designed hand-written programming problems for Python, each of which includes an average of 7.7 test cases per problem.

\item \textbf{CodeHarmony~\cite{Flab-Pruner}:} this dataset contains 16,153 high quality Python code samples synthesized by  LLMs, which is extracted from existing open-source datasets, such as the Evol dataset and OSS dataset. Each problem consists of 3 test cases. In our study, we choose CodeHarmony's test set containing 153 samples.

\item \textbf{MBPP~\cite{austin2021program}:} This benchmark consists of around 1,000 crowd-sourced Python programming problems, designed to be solvable by entry level programmers, covering programming fundamentals, standard library functionality, and so on. Each problem consists of 3 automated test cases. In our study, we choose MBPP's test set containing 500 samples.

\item \textbf{Subtle~\cite{xia2024top}:} this dataset contains 100 programming problems for Python based on HumanEval, which makes a subtle and minor change to the original problem such as inverting or replacing a requirement. Each problem consists of an average of 10.3 test cases per problem.

\item \textbf{Creative~\cite{xia2024top}:} this dataset contains 100 programming problems for Python based on HumanEval, which generates a more creative problem compared to the original through the use of stories or uncommon narratives. Each problem consists of an average of 43.1 test cases per problem.

\item \textbf{BigCodeBench~\cite{zhuo2024bigcodebench}:} A set of 1,140 programming problems for Python which evaluates LLMs with practical and challenging programming tasks. Each problem consists of an average of 5.6 test cases. In our study, we choose BigCodeBench's hard set containing 148 samples.
\end{itemize}

\begin{figure}[t]
\centering
\includegraphics[width=1\textwidth]{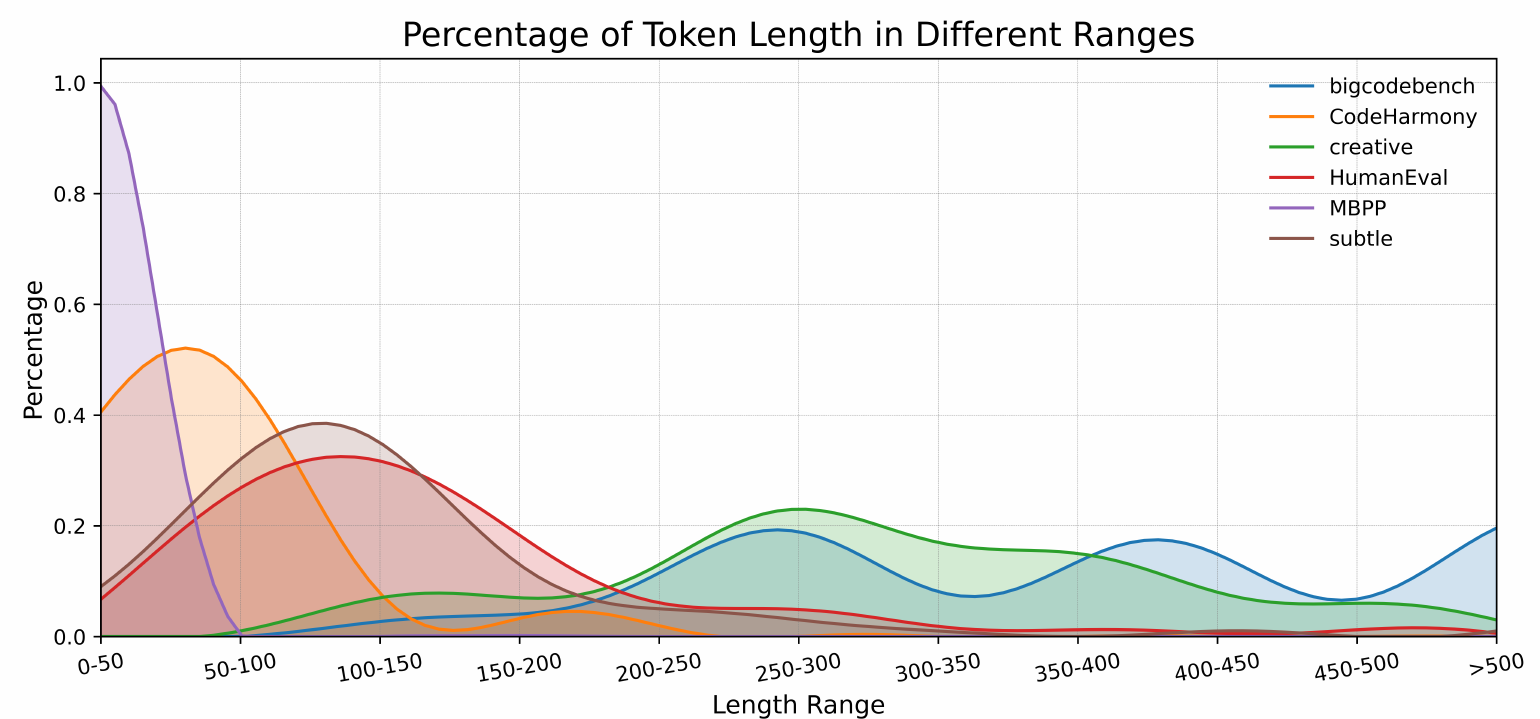}
\caption{The length distribution of DocString in these datasets. Notice for each dataset, the percentage represents the proportion of samples falling within a specific length range relative to the total number of samples. The length range indicates the number of tokens in the DocString after tokenization.}
 \label{fig:dataset}
\vspace{-0.3cm}
\end{figure}

Fig.~\ref{fig:dataset} shows the length distribution of DocString in these datasets. 
%The docstrings in MBPP are shorter, with the lengths mainly centered within 100.
%CodeHarmony, HumanEval, and subtle are more moderate in length, mainly within 300.
%The lengths of creative and BigCodeBench are longer, centered around 300 or more.
The different lengths distribution and diversity in these datasets ensures that our findings are applicable to a broad range of code generation scenarios, from simple script generation to complex software development tasks.

\smallskip
\noindent{\bf Study Models.} We select a diverse set of LLMs that are representative of the current state-of-the-art in code generation, %. The models selected for this empirical study 
which include DeepSeekCoder-1.3b, DeepSeekCoder-6.7b~\cite{guo2024deepseek}, CodeQwen1.5~\cite{bai2023qwen}, CodeGeeX4~\cite{zheng2023codegeex} and Llama3.1~\cite{dubey2024llama}. 

\begin{itemize}
\item \textbf{DeepSeekCoder:} The 1.3b and 6.7b versions of DeepSeekCoder are trained on an extensive dataset, consisting of 87\% code and 13\% natural language in both English and Chinese. These models undergo pre-training on a project-level code corpus with a window size of 16K tokens, enhanced by a fill-in-the-blank task designed to bolster their project-level code completion and infilling capabilities.

\item \textbf{CodeQwen1.5:} This model boasts 7.3 billion parameters and supports an impressive 92 programming languages, managing contexts up to 64K tokens with ease. It has demonstrated remarkable proficiency in code generation, long sequence modeling, code modification, and SQL capabilities.

\item \textbf{CodeGeeX4:} With 9.4 billion parameters at its disposal, CodeGeeX4 is tailored for a myriad of AI software development scenarios. Its strengths lie in code completion, interpretation, web search, function calling, and repository-level question-and-answer interactions.

\item \textbf{Llama3.1:} The 8.0 billion parameter version of Llama3.1 is an open-source powerhouse, trained on a grand scale with the ability to handle context lengths up to 128K tokens. 

\end{itemize}

Among these, DeepSeekCoder-1.3b, DeepSeekCoder-6.7b, CodeQwen1.5 and CodeGeeX4 are specialized for coding tasks, while Llama3.1, being a general-purpose model, has %also 
proven %its mettle in code-related tasks, showcasing 
its versatility and potential for applications in diverse coding scenarios. %\tl{to be improved later.}

% \item \textbf{GPT-4o:} It serves as a benchmark for state-of-the-art performance in code generation. As a closed-source model, it provides a point of comparison for the capabilities and efficiency of the open-source models included in our study.
\smallskip
\noindent{\bf Existing Prompt Compression Methods.}
%In the selection of existing methods, 
We explore two state-of-the-art  prompt compression methods: Selective\_Context~\cite{li2023compressing} and LLMLingua2~\cite{pan-etal-2024-llmlingua}. 

\textbf{Selective\_Context} employs a base language model (such as GPT-2) to compute the self-information of lexical units such as sentences, phrases or tokens, which is then used to evaluate the informativeness. 
Specifically, it operates at the noun phrase level by first using NLTK for lexical analysis, then applying GPT tokenizer for subword tokenization, and finally merging tokens into noun phrases using spacy, which explains why some compressed tokens appear as subwords (e.g., "tu" for "tuples") in Fig.~\ref{fig:example}.

\textbf{LLMLingua2} is a task-agnostic prompt compression technique, i.e., it does not depend on the information entropy of the prompt. It utilizes data distillation to create an effective prompt compression dataset and trains a classification model as a way to select tokens that can be compressed.
Unlike Selective\_Context, LLMLingua2 operates at the complete word level, using a BERT classifier to identify redundant tokens.

\smallskip
\noindent{\bf Evaluation Metrics.}
%In the realm of 
For code generation %, evaluating the performance of models is crucial. Two key metrics that are often utilized are 
Pass@1 and Compression Ratio are considered. Pass@1 indicates the percentage of tasks for which the model produces a code snippet that successfully passes all the associated test cases on the first try. This metric is a direct reflection of the model's accuracy and reliability in code generation scenarios.
Compression Ratio %In the context of prompt compression, the compression ratio is a critical metric that 
quantifies the efficiency of the compression technique. 
It is defined as the percentage reduction in the length of the prompt after compression relative to the original length, i.e., 
$1 - \frac{\|D'\|}{\|D\|}$, where $D$ is the original DocString and $D'$ is the compressed one.
A higher compression ratio indicates that more of the original prompt has been removed, which can lead to improvements in inference speed and reduced computational costs.

%However, it is also essential to ensure that the compression does not compromise the quality of the generated code. The ideal compression method would achieve a high compression ratio while maintaining or even enhancing the performance of the model, as indicated by metrics like Pass@1.

\smallskip 
\noindent{\bf Implementation.} 
All LLMs and their corresponding tokenizers are loaded from the official Huggingface repository, ensuring that the  up-to-date and optimized versions are used in our experiments. 
To ensure a fair comparison, we maintain the hyperparameters of all models consistent throughout our study. 
For the backend service, we utilize the VLLM~\cite{kwon2023efficient}, which provides a unified interface for model inference. 
In addition, we employ Greedy Search as the inference strategy for all models. This method is chosen for its simplicity and effectiveness in generating sequential text, which is a common requirement in code generation tasks. 
Finally, the maximum output length for all models is set to 512 tokens. 

%%%%%%%%%%%%%%%%%%%%%%%%%%%%%%%%%%%%%%%%%%%%%%%%%%%%

\subsection{Empirical Findings}

\begin{table*}[t]
\centering
\caption{Pass@1 Comparison of Existing Compression Methods during Different Ratios}
\resizebox{\textwidth}{!}{
\begin{tabular}{cc||c||cccc||cccc||c}
\toprule
\multirow{2}{*}{\textbf{Model}} & \multirow{2}{*}{\textbf{Dataset}} &  & \multicolumn{4}{c||}{\textbf{SelectiveContext} (\%)} & \multicolumn{4}{c||}{\textbf{LLMLingua2} (\%)} & \\ 
\cmidrule(lr){4-7} \cmidrule(lr){8-11}
& & 0 & 10 & 20 & 30 & 40 & 10 & 20 & 30 & 40 & 100\\
\midrule
\multirow{7}{*}{DS-1.3B} 
 & HumanEval & 63.42 & 54.88 & 50.00 & 44.51 & 51.22 & 53.05 & 51.22 & 42.68 & 37.20 & 21.95 \\
 & CodeHarmony & 59.48 & 58.82 & 56.21 & 54.90 & 52.29 & 58.82 & 58.82 & 54.25 & 50.33 & 39.22 \\
 & MBPP & 36.40 & 36.60 & 35.40 & 35.60 & 34.20 & 35.60 & 36.80 & 34.20 & 31.80 & 25.00 \\
 & Subtle & 53.00 & 53.00 & 44.00 & 45.00 & 40.00 & 50.00 & 46.00 & 44.00 & 32.00 & 15.00 \\
 & Creative & 23.00 & 21.00 & 20.00 & 17.00 & 14.00 & 20.00 & 22.00 & 17.00 & 14.00 & 2.00\\
 & BigCodeBench & 6.10 & 4.10 & 2.00 & 0.00 & 2.00& 4.10 & 4.10 & 2.00 & 2.00 & 0.70 \\
 \cmidrule(lr){2-12}
 & Avg. & 40.23 & 38.07 & 34.60 & 32.84 & 32.29 & 36.93 & 36.49 & 32.36 & 27.89 & 17.20 \\
\midrule
\multirow{7}{*}{DS-6.7B} 
 & HumanEval & 71.95 & 70.12 & 64.63 & 61.59 & 57.32 & 73.17 & 62.20 & 54.27 & 46.34 & 25.61 \\
 & CodeHarmony & 64.36 & 66.67 & 67.97 & 67.32 & 59.48 & 67.32 & 66.67 & 67.97 & 63.40 & 45.75\\
 & MBPP & 46.20 & 47.60 & 46.60 & 46.20 & 45.00 & 45.80 & 44.80 & 46.20 & 42.00 & 30.60\\
 & Subtle & 61.00 & 58.00 & 55.00 & 55.00 & 47.00 & 58.00 & 60.00 & 45.00 & 39.00 & 13.00\\
 & Creative & 34.00 & 34.00 & 31.00 & 31.00 & 23.00 & 33.00 & 37.00 & 31.00 & 27.00 & 4.00\\
 & BigCodeBench & 12.20 & 8.80 & 4.70 & 6.10 & 6.10& 12.20 & 9.50 & 6.10 & 5.40 &  0.00\\
 \cmidrule(lr){2-12}
 & Avg. & 48.29 & 47.53 & 44.98 & 44.54 & 39.65 & 48.25 & 46.70 & 41.76 & 37.19 & 19.83 \\
\midrule
\multirow{7}{*}{CQ-7.3B} 
 & HumanEval & 77.44 & 76.22 & 73.78 & 70.12 & 67.07 & 82.32 & 73.78 & 65.24 & 57.32 & 31.10 \\
 & CodeHarmony & 60.78 & 60.78 & 58.82 & 59.48 & 58.82 & 64.05 & 63.40 & 62.75 & 58.82 & 49.67 \\
 & MBPP & 59.00 & 58.20 & 57.80 & 57.40 & 55.60 & 58.20 & 56.20 & 53.00 & 51.60 & 35.60 \\
 & Subtle & 63.00 & 54.00 & 54.00 & 52.00 & 51.00 & 65.00 & 57.00 & 55.00 & 48.00 & 14.00 \\
 & Creative & 38.00 & 32.00 & 36.00 & 33.00 & 23.00 & 34.00 & 33.00 & 32.00 & 25.00 & 6.00 \\
 & BigCodeBench & 13.50 & 8.80 & 6.10 & 6.10 & 5.40& 9.50 & 8.80 & 6.10 & 7.40 & 0.00 \\
 \cmidrule(lr){2-12}
 & Avg. & 51.95 & 48.33 & 47.75 & 46.35 & 43.48 & 52.18 & 48.70 & 45.68 & 41.36 & 22.73 \\
\midrule
\multirow{7}{*}{CG-9.4B} 
 & HumanEval & 60.37 & 62.20 & 54.88 & 56.10 & 49.39 & 62.80 & 61.59 & 52.44 & 37.20 & 20.73\\
 & CodeHarmony & 64.71 & 59.48 & 60.78 & 60.78 & 58.17 & 61.44 & 63.40 & 57.52 & 57.52 & 39.22\\
 & MBPP & 46.60 & 45.80 & 43.40 & 44.00 & 42.40 & 45.20 & 45.60 & 42.80 & 40.80 & 29.60\\
 & Subtle & 64.00 & 57.00 & 53.00 & 52.00 & 46.00 & 66.00 & 59.00 & 51.00 & 40.00 & 15.00\\
 & Creative & 36.00 & 33.00 & 31.00 & 32.00 & 28.00 & 36.00 & 39.00 & 32.00 & 21.00 & 3.00\\
 & BigCodeBench & 14.20 & 4.10 & 4.70 & 8.10 & 6.10& 9.50 & 7.40 & 8.80 & 4.70 & 0.00\\
 \cmidrule(lr){2-12}
 & Avg. &  47.65 & 43.60 & 41.29 & 42.16 & 38.34 & 46.82 & 46.00 & 40.76 & 33.54 & 17.93 \\
\midrule
\multirow{7}{*}{LA-8.0B} 
 & HumanEval & 57.32 & 58.54 & 54.88 & 53.05 & 47.56 & 56.71 & 54.27 & 45.12 & 35.37 & 20.73\\
 & CodeHarmony & 59.48 & 58.17 & 56.86 & 58.17 & 58.17 & 62.09 & 62.75 & 62.75 & 54.90 & 43.79\\
 & MBPP & 41.40 & 42.20 & 39.80 & 41.60 & 37.00 & 41.00 & 41.20 & 38.40 & 35.20 & 24.20\\
 & Subtle & 56.00 & 57.00 & 59.00 & 50.00 & 47.00 & 58.00 & 53.00 & 47.00 & 35.00 & 16.00\\
 & Creative & 34.00 & 29.00 & 30.00 & 24.00 & 21.00 & 38.00 & 34.00 & 28.00 & 25.00 & 2.00\\
 & BigCodeBench & 10.80 & 1.40 & 3.40 & 2.70 & 4.70& 8.10 & 7.40 & 5.40 & 4.10 & 0.00\\
 \cmidrule(lr){2-12}
 & Avg. &  43.17 & 41.05 & 40.66 & 38.25 & 35.91 & 43.98 & 42.10 & 37.78 & 31.60 & 17.79 \\
\bottomrule
\end{tabular}
}
\label{tab:em} 
\end{table*}

%\noindent{\bf Results.} 
In Table~\ref{tab:em}, we show the comparison of Pass@1 across different LLMs after applying Selective\_Context and LLMLingua2 compression techniques. 
In particular, %The data in the table details the performance of 
for each model on different datasets, it shows %including 
Pass@1 with the original prompts (0\% compression) as well as the change in Pass@1 with increasing compression (10\%, 20\%, 30\%, 40\%). %In addition, the last row provides the average performance performance over all datasets.

\noindent
(1) \underline{Effectiveness}. As can be seen from the table, with a 10\% compression rate, most models maintain comparable or even improved Pass@1 scores relative to their original performance. Taking the Pass@1 of CodeGeeX4 model on HumanEval dataset as an example, the original Pass@1 is 60.37\%, the Pass@1 even can up to 62.20\% at 10\% compression rate using Selective\_Context. Moreover, the Pass@1 can up to 62.80\% at 10\% compression rate using LLMLingua2. 

% This shows that after compressing away some of the redundant information, the model still understands the task requirements better and generates correct code. This verifies that there is indeed redundant information in DocString that can be compressed without affecting the correctness of the generated code too much.

However, this improvement is not universal across all datasets. Notably, on the BigCodeBench dataset, most models show a decrease in Pass@1 scores after compression. 
We attribute this phenomenon to the inherent complexity and diversity of DocStrings across different datasets.

These mixed results suggest that, while some DocStrings contain redundant information that can be safely compressed without compromising performance, the effectiveness of compression is highly dependent on the dataset characteristics.
This observation highlights the importance of considering dataset-specific features when applying compression techniques.

\noindent
(2) \underline{Limitations}. As the compression rate increases further, the Pass@1 degradation trend becomes obvious. For example, the Pass@1 of the CodeGeeX4 model on the HumanEval dataset decreases to 49.39\% at a compression rate of 40\% using Selective\_Context, and to 37.20\% at a compression rate of 40\% using LLMLingua2. 
This suggests that existing compression methods, while removing more information, may also remove semantic information that is critical for the model to generate correct code. 

\noindent
(3) \underline{Code generation after removing DocString}. Interestingly, even in extreme cases, such as in the MBPP dataset, at 100\% compression (i.e., DocString is completely removed), the models are still able to generate a certain percentage of correct code. 
This may indicate that in addition to DocString, other elements in the function signature, such as method name, carry important semantic information that is sufficient to guide the model in generating code that functional correct~\cite{ding2024code, yang2024important}. 
This phenomenon is explored further in Section~\ref{sec:discuss}, as it relates to the deeper mechanisms of how the model utilizes different types of cue information to generate code.

\begin{table}[t]
\caption{The top-10 most frequently removed tokens in each dataset}
\centering
\resizebox{\textwidth}{!}{
\begin{tabular}{c|llllllllll}
\toprule
\textbf{DataSet} & \textbf{Rank-1} & \textbf{Rank-2} & \textbf{Rank-3} & \textbf{Rank-4 }& \textbf{Rank-5} & \textbf{Rank-6} & \textbf{Rank-7} & \textbf{Rank-8} & \textbf{Rank-9} & \textbf{Rank-10}\\
\midrule 
\multicolumn{11}{c}{Selective\_Context}    \\
\midrule
HE & of & 2, & Output: & $\Rightarrow$ & to & than & [1, & 1, & [5, & 3,\\
CH & of & to & with & on & Additionally, & returns & otherwise. & by & be & string.\\
MBPP & of & to & not. & list. & string. & array. & tuples. & in & from & number.\\
Subtle & of & 2, & order. & descending & than & to & 3, & [5, & order & by\\
Creative & 3: & to & Output: & on & than & 2: & 2, & ascending & 4, & (1,\\
BCB & be & on & of & Notes: & not & ValueError: & which & [1, & matplotlib. & 2,\\
\midrule 
\multicolumn{11}{c}{LLMLingua2}  \\  
\midrule
HE & the & that & a & and & are & The & an & You & of & is\\
CH & the & a & The & are & that & and & an & This & where & which\\
MBPP & the & a & that & to & an & are & and & is & of & which\\
Subtle & the & a & that & and & are & an & The & of & is & any\\
Creative & The & that & the & an & a & are & and & there & which & where\\
BCB & The & a & the & that & an & then & A & and & are & there\\
\bottomrule
\end{tabular}
}
\label{tab:top10}
\end{table}

%%%%%%%%%%%%%%%%%%%%%%%%%%%%%
\smallskip
\noindent{\bf Compressed Token Analysis.} 
In addition to analyzing the compression ratio of prompt compression methods, %it is insightful to 
we examine the specific tokens that are being removed. % by these methods to enhance our understanding of prompt compression in the context of code generation tasks. 
By reviewing the most frequent tokens that are compressed away, we can gain insights into the types of information that are considered redundant or less critical by the compression algorithms.

To this end, we extract the top-10 most frequently removed tokens with the highest frequency of occurrence for the two methods on the 10\% compression ratio on the six datasets, respectively, and the results are shown in Table~\ref{tab:top10}. We can find that the comparison between Selective\_Context and LLMLingua2 reveals differences in the tokens each method targets for compression. This variation could be due to the distinct algorithms and criteria each method uses to evaluate the importance of tokens.

Specifically, Selective\_Context, which is based on information entropy, focuses more on the semantic coherence of the text. It tends to prioritize the retention of tokens that contribute to the logical flow and understanding of the code, thus it might be less likely to compress away tokens that are crucial for the structure or functionality of the code, even if they are not the most frequent.

On the other hand, LLMLingua2, which utilizes a pre-trained classifier, is more attuned to the significance of information. It is particularly sensitive to articles and prepositions such as "the" and "a," which are common in English but may not carry substantial meaning in the context of code. This method might prioritize compressing these tokens as they are less likely to affect the overall functionality or readability of the code.

The divergence in the tokens targeted by each method underscores the importance of the underlying algorithm's design in determining what constitutes "redundant" information. 
It also highlights the need for a nuanced approach to prompt compression, where the method must balance the reduction of redundancy with the preservation of critical information necessary for code comprehension and execution.

%% file: sections/4approach.tex
\section{Approach}
\label{sec:method}

\begin{figure}[t]
\centering
\includegraphics[width=1\textwidth]{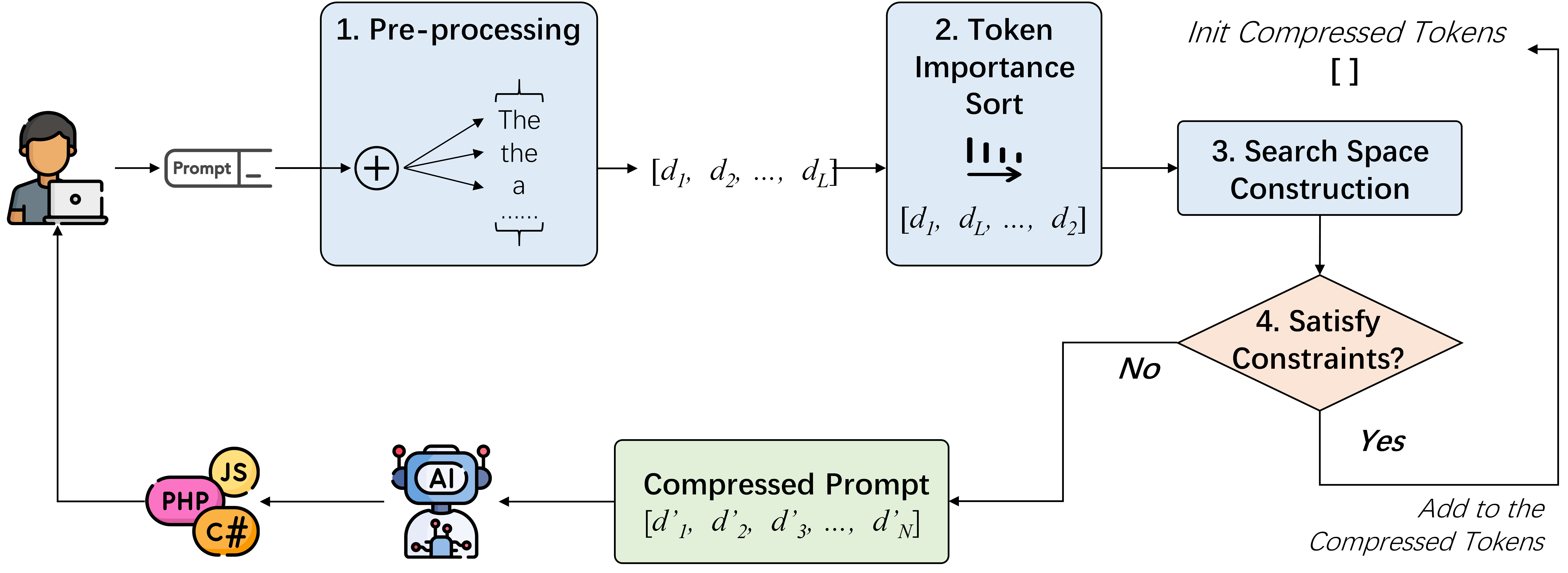}
\caption{Framework of {\tool}.}
 \label{fig:method}
\vspace{-0.3cm}
\end{figure}

Building upon insights from the empirical research, we propose a novel compression technique {\tool}, which is shown in Fig.~\ref{fig:method}. 
%The primary objective of {\tool} is to compress DocStrings in code prompts while preserving their semantic integrity, thereby optimizing the efficiency of LLMs in code generation tasks. 
%The methodology 
Our approach consists of four main stages: Step 1: Pre-processing, Step 2: Token Importance Sort, Step 3: Search Space Construction, and Step 4: Constraint Definition.

% To optimize the consumption of computational resources and time, we select CodeGPT-py-adapted~\cite{lu1codexglue} as the foundational language model for assessing token significance and formulating subsequent constraints. 
% This choice is justified by the model's relatively modest parameter count and minimal GPU memory requirements, which facilitate efficient computations. 
% \tl{what is the point of this remark?} Furthermore, as demonstrated in LLMLingua~\cite{jiang2023llmlingua}, smaller language models such as GPT2-small can maintain comprehension of compressed prompts.

%In the following subsections, we first given the problem definition and then detail each step of the methodology.

%\subsection{Problem Definition}
Given a prompt consisting of a signature $T$ and a DocString $D$, our goal is to compress the DocString into $D^{\prime}$ while preserving the overall semantics and effectiveness in guiding LLMs for code generation.

We formalize this as an optimization problem aiming to find the optimal compressed DocString $D^{\prime}$ that satisfies semantic similarity constraints $\mathcal{C}$ (which we discuss in Section 4.4). The optimization task is defined as:
$$\begin{aligned}
&\underset{D^{\prime}}{\operatorname*{minimize}}&&\|D^{\prime}\|\\
&\mathrm{subject~to}&& \mathcal{C} ,\text{and}~D^{\prime}\sqsubseteq D,
\end{aligned}$$
where $\|D^{\prime}\|$ denotes the length (number of tokens) of the compressed DocString $D^{\prime}$, and $D^{\prime}\sqsubseteq D$ asserts that $D^{\prime}$ is a \emph{subsequence} of the original DocString $D$. 

Specifically, in our approach, we opt for BPE (Byte Pair Encoding) tokenizer's subword level, which aligns with the native tokenization of most modern LLMs while providing fine-grained compression control, and eliminates the need for additional token conversion steps, reducing computational overhead. 
Furthermore, subword tokenization offers finer-grained control over compression. 
For instance, when compressing technical terms or compound words in DocStrings, subword-level operations can preserve critical parts while removing less essential components.
% \tl{this formalisation is very messy, }
% We seek a compressed sequence $\mathcal{D}'$ that adheres to a constraint $\mathcal{C}$ ensuring minimal semantic distortion:
% \[
% \mathcal{D}' \subseteq \mathcal{D} \quad \text{and} \quad \text{CosineSimilarity}(z(S, D), z(S, D')) \geq \tau
% \]
% where $\tau$ is a predefined similarity threshold and $z(S, D)$ represents the semantic vector of the sequence derived from the hidden layers of the base language model.

\subsection{Pre-processing}

Inspired by the previous study~\cite{wang2023recode}, %the addition of line breaks and tabs does not significantly affect the performance of the code generation model, so 
we posit that the removal of line breaks and tabs from  DocString does not significantly affect Pass@1 of the code generation model. (An analysis will be discussed in Section~\ref{sec:discuss}.)

Moreover, upon examining the tokens that frequently undergo compression, as indicated in Table~\ref{tab:top10}, we  observe that the tokens compressed by LLMLingua2 exhibit regular patterns (such as 'the', 'a'), and the model's Pass@1 score remains largely intact when compression is applied at a rate of 10\%. 
This insight suggests that the compression of these tokens does not adversely affect model's Pass@1 score, leading us to categorize them as stop-words.

To minimize semantic distortion in the compressed DocString, we calculate the semantic similarity between the original and the compressed versions for each stop word we attempt to compress. This ensures that the meaning of the DocString remains closely aligned with the original, even after compression.
% we compute the semantic similarity between the original and compressed DocStrings for each stop-word undergoing compression \tl{hard to read}. 
Specifically, we choose the CodeGPT-py-adapted~\cite{lu1codexglue} and adopt the methodology proposed by OpenAI~\cite{neelakantan2022text} for extracting semantic vectors corresponding to the text, followed by the computation of semantic similarity using cosine similarity. 
Compression is performed only for those tokens where the similarity metric exceeds a threshold of 0.999, ensuring the preservation of DocString's semantic integrity.

%%%%%%%%%%%%%%%%%%%%%%%%%%%%%%%%%%%%%%%%%%%%%%%%%%

\subsection{Token Importance Sort}

Following the preprocessing phase, we obtain the token sequence $D = [d_{1}, \ldots, d_{L}]$. To determine the importance of each token in this sequence, we draw upon principles from information theory. 
Specifically, the significance of each token $d_t$ is quantified by its contribution to the conditional entropy of the sequence, which is quantified by its negative log-probability given its preceding context: 
% \tl{how do you get the distribution P?}
\[
\text{Information}(d_{t}) = -\log P(d_{t} \mid d_{1}, d_{2}, \cdots, d_{t-1})
\] 
This measure reflects how informative the token $d_t$ is, given the tokens that come before it. 
A higher value indicates that the token carries more information and is thus more important to the sequence's overall meaning.

To calculate the distribution $P$, %and consider the consumption of computational resources and time, 
we also adopt CodeGPT-py-adapted~\cite{lu1codexglue} as the foundational language model for assessing token significance and formulating subsequent constraints. 
This choice is justified by the model's relatively modest parameter count and minimal GPU memory requirements, which facilitate efficient computation. 
Furthermore, as demonstrated in LLMLingua~\cite{jiang2023llmlingua}, smaller language models such as GPT2-small can maintain comprehension of compressed prompts.

The perplexity of the entire DocString $D$ serves as a measure of its overall information content, calculated as the average conditional entropy across all tokens: 
\[
\text{Perplexity}(D) = -\frac{1}{L} \sum_{i=1}^{L} \log P(d_{i} | d_{1}, d_{2}, \cdots, d_{i-1})
\]
This value serves as a measure of the sequence's overall unpredictability or information content.

To elucidate the individual informational contribution of each token, we compute the perplexity of $D^{-i}$, obtained by excluding a specific token $d_i$ from $D$, viz., 
$
D^{-i} = (d_{1}, d_{2}, \ldots, d_{i-1}, d_{i+1}, \ldots, d_{L})
$
The importance of each token $d_i$ is then calculated as the difference between the perplexity of $D$ and that of $D^{-i}$, formulated as  
\[
\text{Importance}(d_i) = \text{Perplexity}(D) - \text{Perplexity}(D^{-i}) %:= \frac{1}{L-1} \sum_{\substack{j=1 \ j \neq i}}^{L} \log P(d_{j} | d_{1}, d_{2}, \ldots, d_{j-1})
\]
%
%By computing $\text{Importance}(d_i)$ for each token $d_i$ in $D$, we generate a set of importance scores:
%$$\{\mathrm{Importance}(d_1),\mathrm{Importance}(d_2),\ldots,\mathrm{Importance}(d_L)\}$$
We then sort the tokens in ascending order based on their importance scores, i.e., $D'=[d_{i_1}, \cdots, d_{i_L}]$
%$\mathcal{I} = [i_1, i_2, \ldots, i_L]$, 
where $i_1, i_2, \ldots, i_L$ represent the index of each token in $D$.

This approach ensures that tokens whose removal leads to a significant increase in the sequence's conditional entropy are identified as highly important. These tokens are critical for maintaining the semantic integrity of the DocString and are therefore prioritized for retention during the compression process.
Conversely, tokens whose removal results in minimal changes to the conditional entropy are considered less important. These tokens are prime candidates for compression, as their absence is unlikely to adversely affect the model's understanding of the DocString's content.

%%%%%%%%%%%%%%%%%%%%%%%%%%%%%%%%%%%%%%%%5

\subsection{Search Space Construction}
After ranking the tokens by their importance scores, we proceed to construct the search space $\mathcal{S}$  %for the compression process. This search space 
which comprises candidate tokens (sequences) that may be  removed. To efficiently explore the search space and avoid suboptimal solutions due to local optima, we employ a Top-N strategy, i.e., %which 
to select  the top N least important tokens $[d_{i_1}, d_{i_2}, \ldots, d_{i_N}]$.
%means selecting the top N least important tokens and 
% can be described as: 
% % \tl{this is a set, so order does not matter?!}
% \[
% \mathcal{S}_N = [d_{i_1}, d_{i_2}, \ldots, d_{i_N}]
% \]
% \tl{still unclear, where is used?}
%
%Moreover, 
Our empirical observations in Table~\ref{tab:em} indicate that, in certain scenarios, a higher compression rate may yield enhanced Pass@1 score. 
Taking the DS-6.7B model as an example, on the CodeHarmony dataset, the Pass@1 of LLMLingua2 under 20\% compression outperforms that under 10\% compression.
%This leads us to hypothesize 
We hypothesize that the simultaneous compression of consecutive tokens could provide unexpected benefits over compressing individual tokens. As a result, %we construct the N-gram strategy %\\tl{before it is top N strategy?}, 
%which means 
for each $1\leq k\leq N$ we consider $k$-gram, i.e., 
\[
\mathcal{G}_k = \{[d_{i_j}, d_{i_{j+1}}, \ldots, d_{i_{j+k-1}}] \mid 1\leq j\leq N-k+1 \} %i_j \in \mathcal{I}, j+k-1 \leq N]
\]
%where $\mathcal{I}$ is defined in the above Section 4.2.
Each $\mathcal{G}_k$ represents the set of all possible consecutive k-grams within the Top-N tokens.

The complete search space $\mathcal{S}$ is the union of all $\mathcal{G}_k$:
\[
\mathcal{S} = \bigcup_{k=1}^N \mathcal{G}_k
\]
This approach %ensures  
considers both individual low-importance tokens and sequences of such tokens, thereby exploring a richer set of compression candidates.

In terms of computational complexity, the number of combinations for each $k$-gram is $N-k+1$, so the size of $\mathcal{S}$ %the total complexity 
is bounded by $\frac{N(N + 1)}{2}$.
%This results in a computational complexity of $O(N^2)$, which is manageable for moderate values of N.

For example, let us assume that $N = 5$ and $\mathcal{I} = [2, 15, 3, 10, 13]$, then the search space can be defined as 
\begin{itemize}
    \item 1-grams: $\mathcal{G}_1 = \{[2], [15], [3], [10], [13]\}$
    \item 2-grams: $\mathcal{G}_2 = \{[2, 15], [15, 3], [3, 10], [10, 13]\}$
    \item 3-grams: $\mathcal{G}_3 = \{[2, 15, 3], [15, 3, 10], [3, 10, 13]\}$
    \item 4-grams: $\mathcal{G}_4 = \{[2, 15, 3, 10], [15, 3, 10, 13]\}$
    \item 5-grams: $\mathcal{G}_5 = \{[2, 15, 3, 10, 13]\}$
\end{itemize}

By including N-gram combinations, we capture the potential benefits of compressing sequences of tokens that may collectively have a low impact on the semantic content but offer greater compression efficiency when removed together. 
This comprehensive search space allows us to explore various compression strategies and select the one that best satisfies our optimization objectives and constraints.

\subsection{Constraint Definition}
In the context of code generation, the probability of generating a sequence of code $Y$ by given a signature $T$ and a DocString $D$ is defined as:
\[
P(Y|T, D) = \prod_{i=1}^{m} P(y_i | T, D, y_1, y_2, \ldots, y_{i-1})
\]
Here, $T$ is considered constant for different compressions, implying that the variability in \( P(Y|T, D) \) is attributed to changes in \( D \). We denote the compressed version of \( D \) as \( D^{\prime} \), and the corresponding probability as:
\[
P(Y|T, D^{\prime}) = \prod_{i=1}^{m} P(y_i | T, D^{\prime}, y_1, y_2, \ldots, y_{i-1})
\]
Our goal is to minimize the difference between $P(Y|T, D)$ and $P(Y|T, D^{\prime})$, ensuring that the compression does not significantly alter the model's output.
\[
P(Y|T, D) \approx P(Y|T, D^{\prime})
\]

However, directly computing the divergence between these two distributions over all possible code sequences $\mathcal{Y}$ is computationally infeasible. 
To address this, we approximate the requirement by focusing on the model's output distribution for the next token.
Given the sequential nature of code generation, where each token depends on the previous tokens, we simplify the comparison to the probability of generating the first token $y_1$:
\[
P(y_1|T, D) \approx P(y_1|T, D^{\prime})
\]

%Based on the knowledge we discussed in Section 2, 
The probability of the first token is given by the softmax function of the model's output:
\[
P(y_1 |T, D) := \text{softmax}(z(T, D))
\]
Thus, the constraint can be expressed as the proximity between the softmax inputs for the original and compressed sequences:
\[
\text{softmax}(z(T, D)) \approx \text{softmax}(z(T, D^{\prime}))
\]

To ensure that the model's behavior remains consistent after compression, we impose a constraint on the similarity between the logits $z(T, D)$ and $z(T, D^{\prime})$. We use the cosine similarity metric to quantify the closeness between these two vectors:
\[
\text{CosineSimilarity}(z(T, D), z(T, D^{\prime})) = \frac{z(T, D) \cdot z(T, D^{\prime})}{\|z(T, D)\| \cdot \|z(T, D^{\prime})\|}
\]
Our constraint is then formalized as:
\[
\text{CosineSimilarity}(z(T, D), z(T, D^{\prime})) \geq \tau
\]
Here, $\tau$ is the hyper-parameter which is set to be 0.999 in our study.

Integrating this constraint into our overall optimization framework, we aim to find the compressed DocString $D^{\prime}$ that minimizes its length while satisfying the semantic similarity constraint:
$$\begin{aligned}
&\underset{D^{\prime}}{\operatorname*{minimize}}&&\|D^{\prime}\|\\
&\mathrm{subject~to}&& D^{\prime}\sqsubseteq D, \ \\
& && \text{CosineSimilarity}(z(T, D), z(T, D^{\prime})) \geq \tau 
\end{aligned}$$

%% file: sections/5results.tex
\section{Evaluation}
\label{sec:evaluation}

We evaluate the effectiveness of our proposed approach 
from three perspectives. %we mainly design the following three research questions (RQs):

% ==================================================================================
\subsection{RQ1: What is the effectiveness of our proposed {\tool} compared to existing prompt compression methods? (Pass@1 Comparison)}

We aim to evaluate the effectiveness of {\tool} in %significantly 
reducing DocString length without compromising the Pass@1 score of code generation models. 
To ensure the robustness of our findings, we also include a closed-source state-of-the-art model \textbf{GPT-4o} in our analysis. Additionally, we introduce the Random method as a baseline. %, which involves nondeterministic token compression. 
The focus of our analysis is on the Pass@1 score and the compression ratio across a spectrum of datasets and models, as presented in Table~\ref{tab:RQ1}. 
%\tl{explain the table a bit, what does - stand for?}

The compression ratios for different datasets are carefully determined through preliminary experiments, considering the following factors:
\begin{itemize}
    \item \textbf{Dataset Characteristics:} Different datasets exhibit varying DocString patterns and information density. For instance, MBPP's DocStrings contain more redundant information, allowing a higher compression ratio (38\%), while CodeHarmony's more concise DocStrings necessitate a lower ratio (25\%).
    \item \textbf{Fair Comparison:} To ensure a fair comparison, we apply the same compression ratio to all methods within each dataset. While different methods might achieve their optimal performance at different ratios, using consistent ratios within datasets allows a direct comparison of compression effectiveness.
\end{itemize}

% 1、性能对比
\begin{table*}[t]
\centering
\caption{Pass@1 Comparison of Different Methods Across Various Datasets and Models}
\resizebox{\textwidth}{!}{
\begin{tabular}{c|c|c|cccccc}
\toprule
\multirow{2}{*}{\textbf{DataSet}} & \multirow{2}{*}{\textbf{Method}} & \multirow{2}{*}{\textbf{Ratio}} & \multicolumn{6}{c}{\textbf{LLMs}} \\
 & & & \textbf{DS-1.3B} & \textbf{DS-6.7B} & \textbf{CQ-7.3B} & \textbf{CG-9.4B} & \textbf{LA-8.0B} & \textbf{GPT-4o}\\ 
\midrule
\multirow{5}{*}{HumanEval} 
 & None & 0 & 63.42 & 71.95 & 77.44 & 60.37 & 57.32 & 85.37\\
 \cmidrule(lr){2-9}
 & Random & 30 & 32.32 & 38.41 & 60.37 & 36.59 & 35.37 & 50.61\\
 & SelectiveContext & 30 & 44.51 & 61.59 & 70.12 & 56.10 & 53.05 & 75.00\\
 & LLMLingua2 & 30 & 42.68 & 54.27 & 65.24 & 52.44 & 45.12 & 67.68\\
 & {\tool} & 30 & \textbf{57.93} & \textbf{72.56} & \textbf{78.66} & \textbf{64.34} & \textbf{56.10} & \textbf{83.54}\\
\midrule
\multirow{5}{*}{CodeHarmony} 
 & None & 0 & 59.48 & 64.36 & 60.78 & 64.71 & 59.48 & 62.09\\
 \cmidrule(lr){2-9}
 & Random & 25 & 49.63 & 60.78 & 51.63 & 52.29 & 56.21 & 54.25\\
 & SelectiveContext & 25 & 54.90 & 67.32 & 56.86 & 59.48 & 59.48 & 61.44\\
 & LLMLingua2 & 25 & 54.25 & \textbf{67.97} & 61.40 & 59.48 & \textbf{62.09} & 61.44\\
 & {\tool} & 25 & \textbf{55.56} & 66.01 & \textbf{63.40} & \textbf{60.78} & 61.44 & \textbf{62.09}\\
\midrule
\multirow{5}{*}{MBPP} 
 & None & 0 & 36.40 & 46.20 & 59.00 & 46.60 & 41.40 & 55.00\\
 \cmidrule(lr){2-9}
 & Random & 38 & 30.20 & 39.40 & 48.20 & 38.60 & 32.00 & 46.20\\
 & SelectiveContext & 38 & 34.60 & \textbf{44.80} & \textbf{55.40} & 42.00 & 37.60 & 50.00\\
 & LLMLingua2 & 38 & 31.00 & 42.00 & 52.40 & 42.20 & 37.20 & 50.00\\
 & {\tool} & 38 & \textbf{36.60} & 43.80 & \textbf{55.40} & \textbf{46.20} & \textbf{41.40} & \textbf{51.40}\\
\midrule
\multirow{5}{*}{Subtle} 
 & None & 0 & 53.00 & 61.00 & 63.00 & 64.00 & 56.00 & 81.00\\
 \cmidrule(lr){2-9}
 & Random & 30 & 28.00 & 35.00 & 40.00 & 35.00 & 34.00 & 42.00\\
 & SelectiveContext & 30 & 45.00 & 55.00 & 52.00 & 52.00 & 50.00 & 64.00\\
 & LLMLingua2 & 30 & 44.00 & 45.00 & 55.00 & 51.00 & 47.00 & 60.00\\
 & {\tool} & 30 & \textbf{54.00} & \textbf{56.00} & \textbf{62.00} & \textbf{66.00} & \textbf{56.00} & \textbf{79.00}\\
\midrule
\multirow{5}{*}{Creative} 
 & None & 0 & 23.00 & 34.00 & 38.00 & 36.00 & 34.00 & 54.00\\
 \cmidrule(lr){2-9}
 & Random & 25 & 9.00 & 16.00 & 18.00 & 21.00 & 21.00 & 29.00\\
 & SelectiveContext & 25 & 19.00 & 28.00 & 29.00 & 32.00 & 28.00 & 49.00\\
 & LLMLingua2 & 25 & \textbf{23.00} & 33.00 & 29.00 & 31.00 & \textbf{35.00} & 51.00\\
 & {\tool} & 25 & \textbf{23.00} & \textbf{36.00} & \textbf{37.00} & \textbf{35.00} & \textbf{35.00} & \textbf{58.00}\\
\midrule
\multirow{5}{*}{BigCodeBench} 
 & None & 0 & 6.10 & 12.20 & 13.50 & 14.20 & 10.80 & 27.70\\
 \cmidrule(lr){2-9}
 & Random & 34 & 0.00 & 5.40 & 4.70 & 5.40 & 1.40 & 12.20\\
 & SelectiveContext & 34 & \textbf{2.00} & \textbf{6.10} & 6.10 & 8.10 & 2.70 & 15.50\\
 & LLMLingua2 & 34 & \textbf{2.00} & 5.40 & 6.10 & 8.80 & 4.10 & 15.50\\
 & {\tool} & 34 & \textbf{2.00} & \textbf{6.10} & \textbf{10.10} & \textbf{9.50} & \textbf{4.70} & \textbf{20.30}\\
\bottomrule
\end{tabular}
}
\label{tab:RQ1} 
\end{table*}

\begin{table*}[t]
\centering
\caption{FLOPs Comparison of Across Various Datasets and Models, the brackets indicate the FLOPs needed for the {\tool} calculation.}
\resizebox{\textwidth}{!}{
\begin{tabular}{c|cc|cc|cc|cc|cc}
\toprule
\multirow{2}{*}{\textbf{DataSet}} & \multicolumn{2}{c|}{\textbf{DS-1.3B}} & \multicolumn{2}{c|}{\textbf{DS-6.7B}} & \multicolumn{2}{c|}{\textbf{CQ-7.3B}} & \multicolumn{2}{c|}{\textbf{CG-9.4B}} & \multicolumn{2}{c}{\textbf{LA-8.0B}} \\ 
& Compress & Raw & Compress & Raw & Compress & Raw & Compress & Raw & Compress & Raw\\
\midrule
HumanEval & 0.28(+0.02) & 0.34 & 1.43(+0.02) & 1.74 & 1.43(+0.02) & 1.79 & 1.70(+0.02) & 2.02 & 1.46(+0.02) & 1.73\\
CodeHarmony & 0.12(+0.01) & 0.15 & 0.62(+0.01) & 0.75 & 0.63(+0.01) & 0.78 & 0.75(+0.01) & 0.88 & 0.65(+0.01) & 0.75\\
MBPP & 0.03(+0.01) & 0.04 & 0.13(+0.01) & 0.22 & 0.14(+0.01) & 0.22 & 0.18(+0.01) & 0.28 & 0.15(+0.01) & 0.24\\
Subtle & 0.25(+0.02) & 0.30 & 1.31(+0.02) & 1.55 & 1.29(+0.02) & 1.59 & 1.53(+0.02) & 1.77 & 1.31(+0.02) & 1.52\\
Creative & 0.65(+0.02) & 0.75 & 3.37(+0.02) & 3.89 & 3.35(+0.02) & 3.90 & 3.97(+0.02) & 4.41 & 3.39(+0.02) & 3.75\\
BigCodeBench & 0.70(+0.05) & 0.85 & 3.59(+0.05) & 4.39 & 3.53(+0.05) & 4.44 & 4.00(+0.05) & 4.67 & 3.41(+0.05) & 3.98\\
\bottomrule
\end{tabular}
}
\label{tab:flops} 
\end{table*}

Our results demonstrate that {\tool} consistently outperforms alternative methods across all datasets and models, particularly at higher compression rates. 
For instance, on the HumanEval dataset, {\tool} achieves the highest Pass@1 score across all models, effectively maintaining or even improving code generation quality despite a 30\% reduction in DocString length. 
This indicates that {\tool} can sustain, and in some cases enhance, model Pass@1 score while significantly compressing input prompts.

Moreover, in several cases, {\tool} surpasses the uncompressed baseline. 
For instance, on the HumanEval dataset, DeepSeekCoder-6.7B, CodeQwen and CodeGeeX4 can even achieve a higher Pass@1 score %than their original Pass@1 value 
when compressed by {\tool}.
This surprising result suggests that {\tool}  not only preserves essential information but may also eliminate redundant or distracting content from the DocStrings, thereby improving model's focus on relevant information during code generation.

Consistently, across other datasets such as CodeHarmony, MBPP, Subtle, Creative and BigCodeBench, {\tool} either matches or exceeds the competing methods, especially under aggressive compression scenarios. The uniform superiority of {\tool} across diverse datasets and models underscores its versatility and robustness. These results suggest that {\tool} effectively balances compression and information retention, making it a potent tool for optimizing prompt inputs in code generation tasks.

Furthermore, the fact that {\tool} maintains a high Pass@1 score across models of varying sizes and architectures indicates its generalizability. It effectively aids both open-source and closed-source models, including state-of-the-art systems like GPT-4o. This broad applicability enhances the practical value of {\tool} in real-world code generation applications.
 
\textbf{Efficiency:} 
To comprehensively evaluate the efficiency of {\tool}, we analyze its performance from three perspectives:

\textbf{(1) GPU Memory Usage:} GPU memory consumption is a critical resource constraint in deploying LLMs for code generation tasks. 
To evaluate the memory efficiency of our approach, we compare the GPU memory usage between using compressed and uncompressed DocStrings.

It is worth noting that, while our method employs CodeGPT-py-adapted as the base model for compression, its memory footprint is negligible compared to the deployment of LLMs. Specifically, CodeGPT-py-adapted has only 124M parameters and requires merely 0.23GB of GPU memory under the torch.bfloat16 precision. This overhead is insignificant when compared to the substantial memory requirements of modern LLMs used for code generation.

\textbf{(2) FLOPs Analysis:} FLOPs (Floating Point Operations) refers to the number of floating-point operations, an indicator used to measure the complexity of a model, reflecting the total number of floating-point operations required during its execution. 
FLOPs is closely related to the input length and structure of the model. 
The purpose of DocString compression is to reduce computational costs and enhance model efficiency. 
Therefore, we utilize the FLOPs metric to evaluate the computational load of different models before and after compression across various datasets.

The results in Table \ref{tab:flops} indicate that the computational load of the models generally decreases after compression.
It can be observed that the FLOPs values for the compressed models (Compress) are lower than those of the original models (Raw) across all datasets. 
For instance, on the HumanEval dataset, the FLOPs decreased from 0.34 to 0.28 after compression, which is a reduction of approximately 17.6\%. On the CodeHarmony dataset, the FLOPs decreased from 0.15 to 0.12 after compression, which is a reduction of about 20\%.
This reduction in computational load due to compression means that under the same hardware conditions, the model can process data more quickly, or process more data in the same amount of time.

\textbf{(3) Inference Time:} End-to-end inference time is a crucial metric for real-world applications, directly impacting user experience and system throughput. 
To assess the practical benefits of our compression approach, we conduct comprehensive timing measurements under controlled experimental conditions using NVIDIA RTX 3090 GPU. 
All models were configured to use the torch.bfloat16 precision with both input and output lengths set to 512 tokens.

The results show that the code generation time varies significantly across different model scales. 
The DeepSeekCoder model with 1.3B parameters requires approximately 9.21 seconds per generation, while larger models around 7B parameters take about 17.77 seconds. 
For comparison, when using the GPT-4 API, the generation time typically ranges from 5 to 10 seconds, though it may  fluctuate depending on the server load.

The time overhead introduced by {\tool}'s compression process is relatively modest, averaging around 2 seconds per DocString. 
This additional preprocessing time proves to be a worthwhile investment. 
In practical applications, the total processing pipeline (compression plus inference) remains highly efficient.

\begin{tcolorbox}[width=1.0\linewidth, title={Summary of RQ1}]
{\tool} consistently outperforms existing compression methods across multiple datasets and models, effectively reducing DocString length without compromising, and sometimes even enhancing code generation Pass@1 score. 
% Its ability to maintain or improve model output quality at higher compression rates highlights its efficacy and robustness, making it a valuable approach for optimizing code generation prompts.
\end{tcolorbox}

\subsection{RQ2: What is the effectiveness of our proposed {\tool} in other program languages? (Generalization Capability)}
\begin{table*}[t]
\centering
\caption{Generalization Capability of {\tool} in Different Programming Languages}
\resizebox{\textwidth}{!}{
\begin{tabular}{c|c|c|cccccc}
\toprule
\multirow{2}{*}{\textbf{DataSet}} & \multirow{2}{*}{\textbf{Method}} & \multirow{2}{*}{\textbf{Ratio}} & \multicolumn{6}{c}{\textbf{LLMs}} \\
 & & & \textbf{DS-1.3B} & \textbf{DS-6.7B} & \textbf{CQ-7.3B} & \textbf{CG-9.4B} & \textbf{LA-8.0B} & \textbf{Avg.}\\ 
\midrule
\multirow{5}{*}{CPP} 
 & - & 0 & 47.56 & 63.41 & 65.85 & 59.76 & 43.90 & 56.10 \\ 
 \cmidrule(lr){3-9}
 & Random & 30 & 26.22 & 37.20 & 41.46 & 34.15 & 29.27 & 33.66\\
 & SelectiveContext & 30 & 39.63 & 52.44 & 57.32 & 43.29 & 37.80 & 46.10 \\
 & LLMLingua2 & 30 & 33.54 & 50.00 & 55.49 & 43.90 & 31.71 & 42.93 \\
 & {\tool} & 30 & \textbf{42.07} & \textbf{58.54} & \textbf{60.37} & \textbf{58.54} & \textbf{39.02} & \textbf{51.71} \\
\midrule
\multirow{5}{*}{Go} 
 & - & 0 & 46.95 & 61.59 & 58.54 & 52.44 & 37.80 & 51.46\\
 \cmidrule(lr){3-9}
 & Random & 30 & 21.95 & 32.32 & 35.37 & 27.44 & 22.56 & 27.93\\
 & SelectiveContext & 30 & \textbf{44.51} & 53.05 & 59.76 & 40.24 & 27.44 & 45.00 \\
 & LLMLingua2 & 30 & 34.76 & 42.07 & 55.49 & 36.59 & 26.22 & 39.02 \\
 & {\tool} & 30 & 39.63 & \textbf{56.10} & \textbf{64.63} & \textbf{46.95} & \textbf{31.10} & \textbf{47.68}\\
\midrule
\multirow{5}{*}{Java} 
 & - & 0 & 57.93 & 59.15 & 77.44 & 60.37 & 57.93 & 62.56 \\
 \cmidrule(lr){3-9}
 & Random & 30 & 37.20 & 44.51 & 54.27 & 37.80 & 31.10 & 40.98 \\
 & SelectiveContext & 30 & \textbf{55.49} & 58.54 & 71.95 & 57.32 & 45.73 & 57.80 \\
 & LLMLingua2 & 30 & 46.95 & 51.83 & 67.68 & 54.27 & 34.15 & 50.98 \\
 & {\tool} & 30 & 53.66 & \textbf{67.68} & \textbf{75.00} & \textbf{60.37} & \textbf{48.17} & \textbf{60.98} \\
\midrule
\multirow{5}{*}{JavaScript} 
 & - & 0 & 57.93 & 64.63 & 71.34 & 59.15 & 40.85 & 58.78 \\
 \cmidrule(lr){3-9}
 & Random & 30 & 27.44 & 43.90 & 49.39 & 35.37 & 26.22 & 36.46 \\
 & SelectiveContext & 30 & 46.34 & 55.49 & 57.32 & 43.29 & \textbf{46.34} & 49.76 \\
 & LLMLingua2 & 30 & 45.12 & 48.17 & \textbf{66.46} & 47.56 & 31.71 &47.80 \\
 & {\tool} & 30 & \textbf{56.10} & \textbf{67.07} & 65.24 & \textbf{49.39} & 41.46 & \textbf{55.85} \\
\bottomrule
\end{tabular}
}
\label{tab:RQ2} 
\end{table*}

Considering that the datasets chosen in RQ1 are all in Python, to explore the generalizability of {\tool}, we select  four other programming languages, i.e., C++, Go, Java, JavaScript in the HumanEval-X~\cite{zheng2023codegeex} dataset for our experiments.

Our choice of HumanEval-X is primarily driven by the status quo. %current landscape of code generation datasets. 
At present, most publicly available code generation datasets for non-Python languages are variants of HumanEval, as it has become a \textit{de facto} benchmark in the field. 
While we acknowledge that using variations of a single dataset might introduce potential biases, HumanEval-X represents one of the few comprehensive, well-validated multilingual code generation benchmarks available to the research community.

Since these four datasets share DocString with HumanEval, we directly migrate the compressed DocString obtained for HumanEval in RQ1 to these datasets.
The detailed experimental results are shown in Table~\ref{tab:RQ2}.

The results %as presented in Table~\ref{tab:RQ2} 
reveal that {\tool} maintains its superior performance across all four programming languages when compared to baseline methods (e.g., Random compression, SelectiveContext and LLMLingua2). 
Despite the structural and syntactical differences among these languages, {\tool} consistently achieves higher performance metrics, demonstrating its generalization ability to retain the critical information in DocStrings. %across varied contexts.

Notably, in the C++ and Java datasets, {\tool} significantly outperformed other methods by maintaining closer alignment with the uncompressed baseline, indicating that its approach to token importance and compression strategy is not tightly coupled to any specific programming language. 
The JavaScript dataset, which tends to have more varied DocString usage, still saw considerable gains with {\tool}, reinforcing its versatility. 
The Go dataset showed a similar trend, where {\tool} consistently outperformed the baselines and adapted well to the more concise and functionally focused nature of Go's syntax.

At the same time, we observe a slight degradation in Pass@1 scores when directly migrating the compressed DocStrings obtained in RQ1 (Python) to other programming languages such as C++, Go, Java and JavaScript. 
While {\tool} consistently outperformed baseline methods, its performance did not always fully match the results achieved with the original uncompressed DocStrings in these new languages.

This performance degradation can be attributed to the fact that the base language model used in our approach, CodeGPT, is specifically trained and optimized for Python. The model's understanding of the syntax, structure and semantics of Python-based code influences how DocStrings are compressed. When these compressed DocStrings are applied to other programming languages, certain language-specific nuances may not be as effectively preserved, leading to a small drop in code generation performance.

Despite this limitation, {\tool} still demonstrates strong generalization capability, retaining the critical information needed for code generation tasks across multiple programming languages. This suggests that while language-specific models might yield the best results when paired with {\tool}, the method remains highly effective even when applied across different languages with minimal adjustments.

\begin{tcolorbox}[width=1.0\linewidth, title={Summary of RQ2}] 
While {\tool} generalizes well across programming languages, a slight performance degradation is observed when migrating compressed DocStrings from Python to other languages. This can be attributed to the base model, CodeGPT, being trained specifically for Python, which may not fully capture the nuances of other languages. 
Nonetheless, the generalization capability of {\tool} was thoroughly demonstrated by its performance across C++, Go, Java and JavaScript in the HumanEval-X dataset. By directly migrating compressed DocStrings from Python datasets, {\tool} consistently outperformed baseline methods, proving its adaptability and robustness across multiple programming languages. 
\end{tcolorbox}

\begin{figure}[ht]
\centering
\includegraphics[width=1\textwidth]{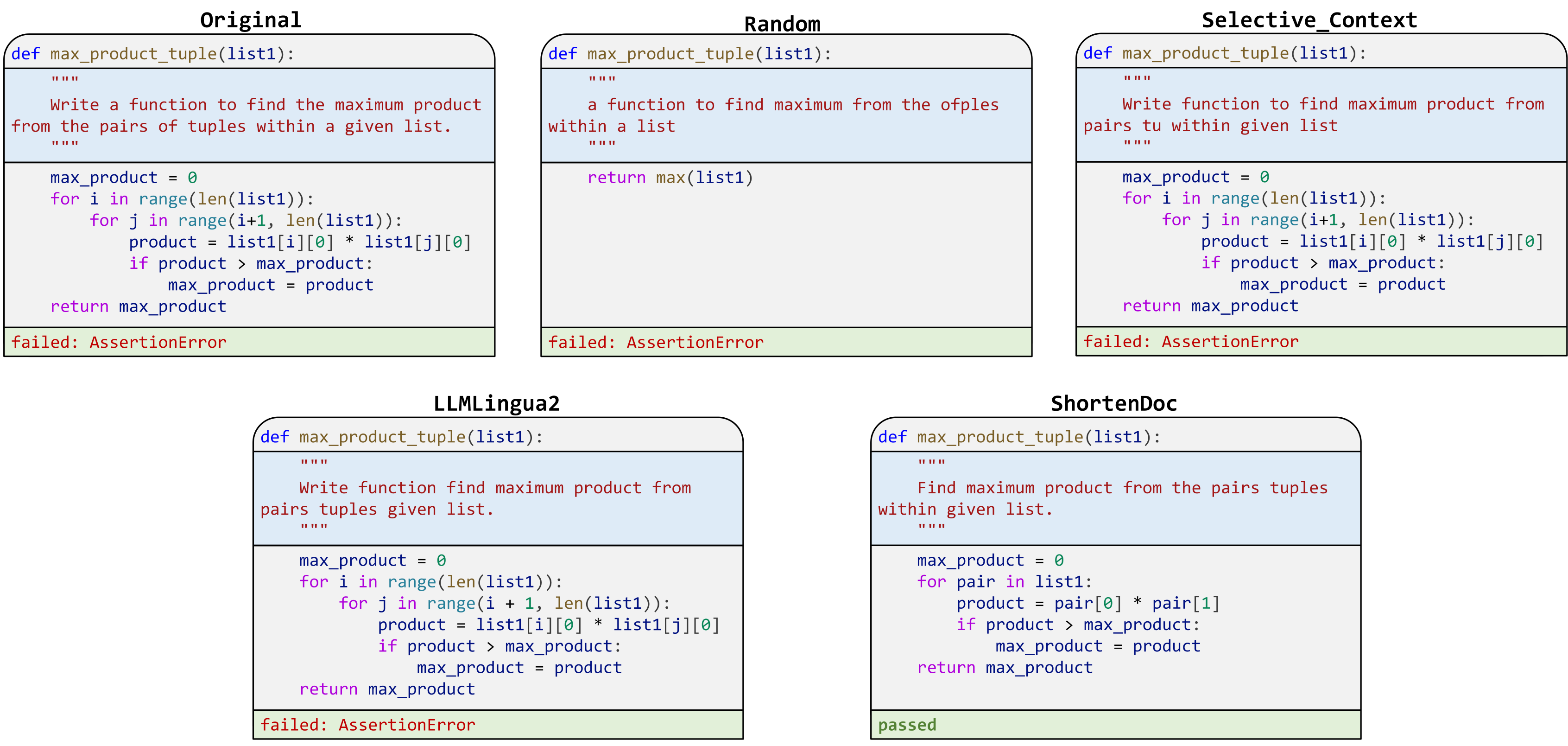}
\caption{\yg{Example of Different DocString Compression Methods}}
 \label{fig:example}
\vspace{-0.3cm}
\end{figure}

%%%%%%%%%%%%%%%%%%%%%%%%%%%%%%%%%%%%%%%%%%%%%

\subsection{RQ3: What is the impact of compressed prompts on developers? (Human Study)}
RQ1 and RQ2 rely on automated evaluation metrics to compare the performance of different methods. 
In RQ3, we aim to gain more insights by conducting a \yg{human study} to assess the impact of DocString compression.

% To begin with, we present a case study to visually contrast the compression effects of our tool with other methods. 
To complement our analysis in RQ1 and RQ2, we first present a single example to provide deeper insights into how different compression methods affect code generation. 
While a single example cannot represent all scenarios, it helps illustrate the typical patterns we have observed across our experiments.
Fig.~\ref{fig:example} %illustrates this by showcasing 
showcases a DocString from the Llama3.1 model within one representative example from the MBPP dataset, showing a DocString processed by different compression methods.
We display the original DocString alongside its compressed versions, processed by four distinct methods. 
The original DocString led to semantically incorrect code generation by the Llama3.1 model. 
The Random compression method resulted in irrelevant code. 
Both Selective\_Context and LLMLingua2 maintained the model's output, yet failed to rectify the semantic inaccuracy. 
In stark contrast, our tool's compression successfully steered the model towards semantically correct code generation.

To further evaluate the quality of the compressed DocStrings, we conducted a human study using three groups of DocStrings compressed by different methods (Selective\_Context, LLMLingua2 and {\tool}). 

By involving human evaluators, we sought to provide a more comprehensive assessment of the compressed DocStrings and to better understand their practical implications. In our human study, we used two key evaluation criteria: Informativeness and Comprehensibility.

\begin{itemize}
\item \textbf{Informativeness}. This criterion measures whether the compressed DocString retains the key information conveyed in the original DocString. A higher score indicates that the essential content is preserved after compression.

\item \textbf{Comprehensibility}. This criterion evaluates how easily a human can understand the compressed DocString. It measures whether the DocString, despite being shortened, remains understandable.
\end{itemize}

\begin{figure}[ht]
\centering
\includegraphics[width=1\textwidth]{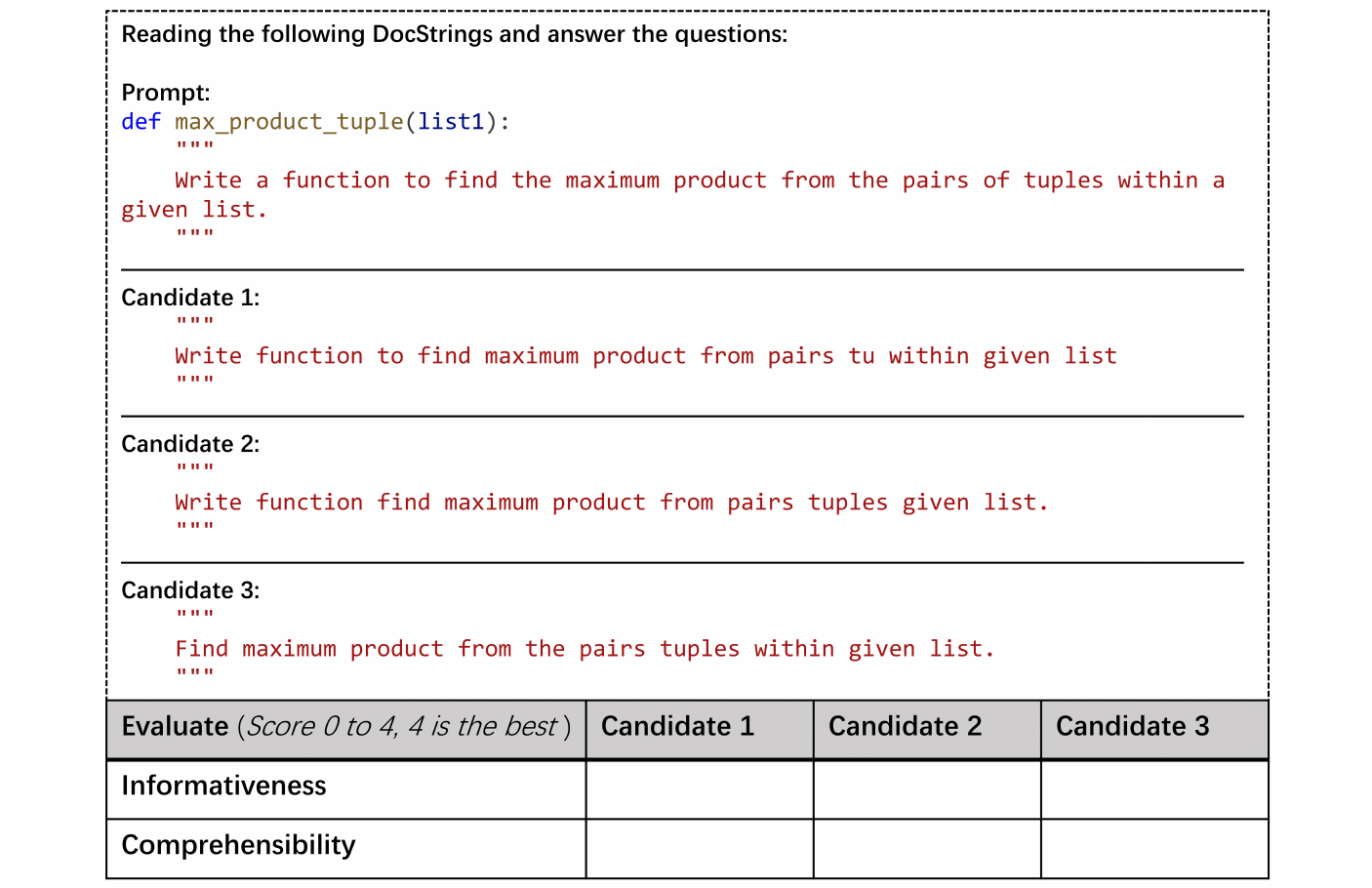}
\caption{A Sample Questionnaire Used in Human Study. The `Prompt' section shows the original uncompressed DocString, while the `Candidates' section presents three compressed versions of the same DocString generated by different methods (SelectiveContext, LLMLingua2 and {\tool}). Evaluators were asked to rate each candidate on Informativeness (preservation of key information) and Comprehensibility (ease of understanding) using a 0-4 scale, without knowing which method produced each compression.}
 \label{fig:human}
\vspace{-0.3cm}
\end{figure}

\yg{To conduct the human study, we recruited three evaluators with a strong background in Python programming (all are PhD candidates). 
We randomly selected 40 samples from each of the six datasets used in the previous experiments, resulting in a total of 240 samples for evaluation.
For each sample, the evaluators assessed the original DocString along with three compressed versions generated by different methods.}

\yg{To facilitate analysis and comparison across different data sources, we organized the samples into six groups according to their source datasets. Each group was evaluated by all three evaluators. The evaluators assessed each DocString anonymously based on two criteria: Informativeness and Comprehensibility. The evaluation scale ranged from 0 to 4 points, with higher scores indicating better quality. The final score for each criterion was calculated as the average of scores from all three evaluators.}

\yg{The questionnaire used for the study is illustrated in Fig.~\ref{fig:human}. To maintain the quality of the human evaluation, the compressed DocStrings were presented in random order, ensuring that the  evaluators were unaware of which method generated the DocString. 
Additionally, the evaluators were allowed to use Internet to look up unfamiliar concepts. 
To prevent fatigue and maintain concentration, we asked each evaluator to complete the assessment of one dataset (40 samples) over two days, with no more than 20 samples evaluated in any half-day session.}

% To conduct the human study, we recruited three evaluators with a strong background in Python programming (all Ph.D. candidates). 
% We selected all \yg{selected} samples from the six datasets used in the previous experiments as the evaluation subjects. 
% \yg{For each dataset, we randomly selected 40 samples for the human study. 
% For each sample, we collected the original DocString along with the three compressed DocStrings generated by three methods.}

% \yg{To ensure comprehensive evaluation, the samples were divided into six groups according to the dataset, and each group was evaluated by the three evaluators. 
% Each DocString was assessed anonymously in terms of both Informativeness and Comprehensibility. The evaluation scale ranged from 0 to 4 points, where a higher score reflected higher quality. The final score for each criterion was the average of the scores from all three evaluators.}

% The questionnaire used for the study is illustrated in the Fig.~\ref{fig:human}. To maintain the quality of the human evaluation, the compressed DocStrings were presented in random order, ensuring that evaluators were unaware of which method was used to generate each DocString. Additionally, the evaluators were allowed to use the internet to look up unfamiliar concepts, ensuring that they had access to relevant information. To prevent evaluator fatigue and maintain focus, we limited each evaluator to assessing no more than 20 samples per half-day.

\begin{table}[t]
  \centering
  \caption{The average score and kappa score (in parentheses) of human study.}
  \resizebox{\textwidth}{!}{
    \begin{tabular}{c|c|cccccc}
    \toprule
    \multirow{2}{*}{\textbf{Aspect}} & \multirow{2}{*}{\textbf{Method}} & \multicolumn{6}{c}{\textbf{Datasets}}\\
    & & \textbf{HumanEval} & \textbf{CodeHarmony} & \textbf{MBPP} & \textbf{Subtle} & \textbf{Creative} & \textbf{BigCodeBench} \\
    \midrule
    \multirow{3}{*}{Informativeness}
    &  Selective\_Context & 2.83 (0.4) & 3.37 (0.5) & \textbf{3.22} (0.2) & 2.85 (0.3) & 3.55 (0.6) & 3.60 (0.5)\\
    &  LLMLingua2 & 2.85 (0.5) & 3.32 (0.4) & 2.83 (0.5) & 2.80 (0.5) & 3.55 (0.5) & 3.35 (0.5)\\
    &  {\tool} & \textbf{3.83} (0.7) & \textbf{3.75} (0.7) & 3.13 (0.5) & \textbf{3.77} (0.7) & \textbf{3.97} (0.9) & \textbf{3.96} (0.9)\\
    \midrule
    \multirow{3}{*}{Comprehensibility}
    &  Selective\_Context & 1.79 (0.2) & 2.44 (0.1) & 2.86 (0.1) & 1.77 (0.2) & 1.87 (0.1) & 1.90 (0.0)\\
    &  LLMLingua2 & 2.30 (0.4) & 2.92 (0.3) & 2.74 (0.3) & 2.24 (0.4) & 2.85 (0.2) & 2.68 (0.2)\\
    &  {\tool} & \textbf{3.48} (0.3) & \textbf{3.50} (0.3) & \textbf{3.61} (0.2) & \textbf{3.33} (0.3) & \textbf{3.67} (0.2) & \textbf{3.66} (0.1)\\
    \bottomrule
    \end{tabular}%
    }
  \label{tab:human}%
\end{table}%

The results presented in Table~\ref{tab:human} from the human study clearly demonstrate that the proposed {\tool} method outperforms the other two methods, Selective\_Context and LLMLingua2, in both Informativeness and Comprehensibility. This superiority indicates that {\tool} is more effective in retaining key information and maintaining the understandability of DocStrings after compression.

Furthermore, to appraise the consistency among the three evaluators, we computed Fleiss' Kappa statistic~\cite{fleiss1971measuring, randolph2005free}. 
Fleiss' Kappa is a generalization of Cohen's Kappa that is used to measure the reliability of categorical data when there are multiple raters. 
It provides a measure of the degree to which the evaluators agree beyond what would be expected by chance. A Kappa value of 1 indicates perfect agreement, while a value of 0 suggests no agreement beyond chance. 
Negative values indicate agreement less than what would be expected by chance. In our study, a Kappa coefficient was calculated to ensure that the evaluation scores were consistent and reliable across the evaluators. 

In terms of Informativeness, the {\tool} method consistently scores higher than the other two methods across most datasets. This suggests that {\tool} is adept at preserving the essential information within the DocStrings even after compression.
In addition, the high Fleiss' Kappa scores, ranging from 0.5 to 0.9, indicate a strong consensus among the evaluators regarding the Informativeness of the compressed DocStrings. 
This high agreement likely stems from the objective nature of this metric, where evaluators can clearly discern whether the compressed DocString retains the necessary information from the original.

In terms of Comprehensibility, {\tool} also shows superiority than the other two methods in Comprehensibility, indicating that the compressed DocStrings remain understandable to human readers.
While the lower Fleiss' Kappa scores for Comprehensibility, ranging from 0.1 to 0.3, suggest that there is less consistency among the evaluators in this metric compared to Informativeness.

% This variability could be attributed to several factors:
The distinct scores in evaluator agreement between Informativeness and Comprehensibility can be attributed to their fundamental differences:
(1) Subjectivity. Comprehensibility is inherently more subjective than Informativeness, as it depends on the evaluator's personal ability to understand the compressed information.
(2) Complexity of Docstrings. Some DocStrings, despite being compressed, may still contain complex information or jargon that is more challenging for some evaluators to grasp.
(3) Diversity in Evaluator Background. The professional background and experience of the evaluators can influence their assessment of DocString comprehensibility. For instance, evaluators with specific domain knowledge might find certain terms or concepts more comprehensible.

\begin{tcolorbox}[width=1.0\linewidth, title={Summary of RQ3}] 
The human study confirms that {\tool} excels in both Informativeness and Comprehensibility compared to SelectiveContext and LLMLingua2. 
The strong performance of {\tool} across both dimensions demonstrates that it not only compresses DocStrings effectively but also preserves critical information and maintains clarity, making it a practical solution for real-world applications. 
\end{tcolorbox}

%% file: sections/6discussion.tex
\section{Discussion}
\label{sec:discuss}
In this section, we delve into the insights gained from the DocString compression techniques and explore the patterns that emerged from our experiments.

\subsection{The Impact of Method Name Quality} 

\begin{table*}[t]
\centering
\caption{Pass@1 comparison within Different Method Name Quality}
\resizebox{\textwidth}{!}{
\begin{tabular}{c|c|cccccc}
    \toprule
    \multirow{2}{*}{\textbf{DataSet}} & \multirow{2}{*}{\textbf{Method}} & \multicolumn{6}{c}{\textbf{LLMs}} \\
     & & \textbf{DS-1.3B} & \textbf{DS-6.7B} & \textbf{CQ-7.3B} & \textbf{CG-9.4B} & \textbf{LA-8.0B} & \textbf{Avg.}\\ 
\midrule
\multirow{3}{*}{HumanEval} 
 & Original & 63.42 & 71.95 & 77.44 & 60.37 & 57.32 & 66.10 \\
 \cmidrule(lr){2-8}
 & {\tool} & \textbf{57.93} & \textbf{72.56} & \textbf{78.66} & \textbf{64.34} & \textbf{56.10} & \textbf{65.92} \\
 & {\tool} w Foo & 50.61 & 71.95 & 68.90 & 54.27 & 53.05 & 59.96\\
\midrule
\multirow{3}{*}{CodeHarmony} 
 & Original & 59.48 & 64.36 & 60.78 & 64.71 & 59.48 & 61.76 \\
 \cmidrule(lr){2-8}
 & {\tool} & \textbf{55.56} & \textbf{66.01} & \textbf{62.75} & \textbf{60.78} & \textbf{61.44} & \textbf{61.31}\\
 & {\tool} w Foo & 48.37 & 60.13 & 58.82 & 51.63 & 54.25 & 54.64\\
\midrule
\multirow{3}{*}{MBPP} 
 & Original & 36.40 & 46.20 & 59.00 & 46.60 & 41.40 & 45.92 \\
 \cmidrule(lr){2-8}
 & {\tool} & \textbf{36.60} & \textbf{43.80} & \textbf{55.40} & \textbf{46.20} & \textbf{41.40} & \textbf{44.68} \\
 & {\tool} w Foo & 30.80 & 38.60 & 47.20 & 37.40 & 37.20 & 38.24 \\
\midrule
\multirow{3}{*}{Subtle} 
 & Original & 53.00 & 61.00 & 63.00 & 64.00 & 56.00 & 59.40 \\
 \cmidrule(lr){2-8}
 & {\tool} & \textbf{54.00} & \textbf{56.00} & \textbf{62.00} & \textbf{66.00} & \textbf{56.00} & \textbf{58.80} \\
 & {\tool} w Foo & 39.00 & 48.00 & 58.00 & 53.00 & 54.00 & 50.40\\
\midrule
\multirow{3}{*}{Creative} 
 & Original & 23.00 & 34.00 & 38.00 & 36.00 & 34.00 & 33.00 \\
 \cmidrule(lr){2-8}
 & {\tool} & \textbf{23.00} & \textbf{36.00} & \textbf{37.00} & \textbf{35.00} & \textbf{35.00} & \textbf{33.20} \\
 & {\tool} w Foo & 22.00 & 33.00 & 39.00 & 36.00 & 25.00 & 31.00\\
\midrule
\multirow{3}{*}{BigCodeBench} 
 & Original & 6.10 & 12.20 & 13.50 & 14.20 & 10.80 & 11.36 \\
 \cmidrule(lr){2-8}
 & {\tool} & 2.00 & 6.10 & 10.10 & 9.50 & 4.70 & 6.48 \\
 & {\tool} w Rename & \textbf{2.00} & \textbf{7.40} & \textbf{10.80} & \textbf{12.20} & \textbf{8.10} & \textbf{8.10} \\
\bottomrule
\end{tabular}
}
\label{tab:func_name} 
\end{table*}
In Section 3, we observed an intriguing phenomenon: models were capable of generating some correct code without relying on DocStrings, indicating that the method signature itself, particularly the method name, carries substantial information. This observation led us to hypothesize that the quality of method names plays a pivotal role in the success of DocString compression.

To empirically validate this hypothesis, we designed a controlled experiment. We systematically replaced high-quality method names with a generic placeholder, such as `foo'~\cite{yang2024important}, across various datasets including HumanEval, CodeHarmony, MBPP, Subtle and Creative. 

Conversely, we enhanced the BigCodeBench dataset by upgrading the low-quality method names to more descriptive and informative alternatives. We observed that all method names in the dataset were generic, lacking any meaningful information, such as `func\_task'. To address this, we manually replaced each sample with high-quality method names that we crafted to provide better context and clarity.

The outcomes of this experiment are encapsulated in Table~\ref{tab:func_name}. The data clearly demonstrate that method names wield a significant influence on the Pass@1 score of DocString compression. Across the board, models exhibited a notable decline in Pass@1 score when high-quality method names were replaced with the generic `foo'. This finding suggests that a portion of the information compressed by {\tool} is redundant with the information already present in the method name.

Conversely, when low-quality method names were enhanced, we observed a consistent uptick in model Pass@1 score. This improvement underscores the notion that enriching the quality of method names not only aids in the comprehension of the code but also bolsters the robustness of {\tool}'s compression capabilities.

In light of these findings, we advocate for a conscientious approach to method name quality during the DocString compression process. By elevating the quality of method names, we can potentially enhance the clarity of the code and the effectiveness of the compression algorithm. This strategy may yield unexpected benefits, such as more efficient code generation and improved maintainability of the codebase.

\subsection{The Impact of DocString Style}

The style of DocStrings is a critical yet often overlooked aspect of code documentation. In our investigation, we explored the impact of DocString style on the performance of our compression tool and the models' ability to generate accurate and understandable code. Our analysis led to several key findings and considerations.

\noindent{\bf Experiments with Newline and Tab Characters.} 
In Section 4, during the preprocessing phase, we made an assumption that the removal of newline characters and tabs would not substantially affect the model's code generation capabilities. This assumption was based on the idea that these characters, while useful for formatting, might not contribute to the semantic content that the model uses to understand the code structure and purpose.
To test this, we conducted controlled experiments where we systematically removed these characters from the DocStrings and compared the model's Pass@1 score against a control group where these elements were retained. 

\noindent{\bf Impact of Stop Words.} 
We also investigated the role of stop words in DocStrings. 
In our study, we define the stop words are commonly considered to be filler words that do not carry much meaning, which are chosen by LLMLingua2 used in a 10\% ratio we discussed in Section 3.
Our experiments involved removing stop words from DocStrings to assess if the model's Pass@1 score would change. The comparison between the Pass@1 score with and without stop words helped us understand the contribution of these words to the overall comprehension of different LLMs.

\noindent{\bf Impact of Redundant Intruction.} 
Our analysis of the MBPP dataset revealed a prevalent pattern of redundant instructions, specifically the phrase "write a function to." 
Intuitively, we suspected that such repetitive and generic instructions might not be beneficial for the model in understanding the specific requirements of the code it is tasked to generate. 
To analyze the impact of these instructions, we designed a set of controlled experiments where we removed these phrases from the DocStrings. 
The comparison of the model's output with and without these redundant instructions allowed us to evaluate their utility in the context of code generation.

\begin{table*}[t]
\centering
\caption{Pass@1 Comparison whitin Different DocString Styles}
\resizebox{\textwidth}{!}{
\begin{tabular}{c|c|cccccc}
    \toprule
    \multirow{2}{*}{\textbf{DataSet}} & \multirow{2}{*}{\textbf{Method}} & \multicolumn{6}{c}{\textbf{LLMs}} \\
     & & \textbf{DS-1.3B} & \textbf{DS-6.7B} & \textbf{CQ-7.3B} & \textbf{CG-9.4B} & \textbf{LA-8.0B} & \textbf{Avg.}\\ 
\midrule
\multirow{3}{*}{HumanEval} 
 & Original & 63.42 & 71.95 & 77.44 & 60.37 & 57.32 & 66.10 \\
 \cmidrule(lr){2-8}
 & Remove New Line \& Tab & 57.32 & 76.83 & 83.84 & 66.46 & 60.37 & 68.96 \\
 & Remove Stop Words & 54.88 & 66.46 & 76.83 & 56.10 & 56.10 & 62.07 \\
\midrule
\multirow{3}{*}{CodeHarmony} 
 & Original & 59.48 & 64.36 & 60.78 & 64.71 & 59.48 & 61.76  \\
 \cmidrule(lr){2-8}
 & Remove New Line \& Tab & 60.78 & 67.32 & 62.75 & 61.44 & 61.44 & 62.75 \\
 & Remove Stop Words & 57.52 & 67.32 & 60.14 & 60.78 & 58.17 & 60.79 \\
\midrule
\multirow{3}{*}{MBPP} 
 & Original & 36.40 & 46.20 & 59.00 & 46.60 & 41.40 & 45.92 \\
 \cmidrule(lr){2-8}
 & Remove New Line \& Tab & 37.80 & 47.20 & 59.00 & 47.80 & 43.00 & 46.96 \\
 & Remove Stop Words & 34.00 & 45.80 & 54.40 & 44.20 & 41.20 & 43.92 \\
\midrule
\multirow{3}{*}{Subtle} 
 & Original & 53.00 & 61.00 & 63.00 & 64.00 & 56.00 & 59.40 \\
 \cmidrule(lr){2-8}
 & Remove New Line \& Tab & 54.00 & 58.00 & 62.00 & 68.00 & 60.00 & 60.40 \\
 & Remove Stop Words & 46.00 & 56.00 & 61.00 & 62.00 & 55.00 & 56.00 \\
\midrule
\multirow{3}{*}{Creative} 
 & Original & 23.00 & 34.00 & 38.00 & 36.00 & 34.00 & 33.00 \\
 \cmidrule(lr){2-8}
 & Remove New Line \& Tab & 22.00 & 31.00 & 33.00 & 34.00 & 30.00 & 30.00 \\
 & Remove Stop Words & 20.00 & 29.00 & 29.00 & 30.00 & 31.00 & 27.80 \\
\midrule
\multirow{3}{*}{BigCodeBench} 
 & Original & 6.10 & 12.20 & 13.50 & 14.20 & 10.80 & 11.36 \\
 \cmidrule(lr){2-8}
 & Remove New Line \& Tab & 1.40 & 8.80 & 12.20 & 7.40 & 6.80 & 7.32 \\
 & Remove Stop Words & 2.0 & 8.10 & 9.50 & 10.80 & 6.10 & 7.30 \\
\bottomrule
\end{tabular}
}
\label{tab:style} 
\end{table*}

\begin{table*}[t]
\centering
\caption{Pass@1 comparison of Redundant Intruction}
\resizebox{\textwidth}{!}{
\begin{tabular}{c|c|cccccc}
    \toprule
    \multirow{2}{*}{\textbf{DataSet}} & \multirow{2}{*}{\textbf{Method}} & \multicolumn{6}{c}{\textbf{LLMs}} \\
     & & \textbf{DS-1.3B} & \textbf{DS-6.7B} & \textbf{CQ-7.3B} & \textbf{CG-9.4B} & \textbf{LA-8.0B} & \textbf{Avg.}\\ 
\midrule
\multirow{3}{*}{MBPP} 
 & Original & 36.40 & 46.20 & 59.00 & 46.60 & 41.40 & 45.92 \\
 & w/o `Write a python function to' & 37.80 & 45.60 & 57.40 & 47.60 & 42.80 & 46.24 \\
\bottomrule
\end{tabular}
}
\label{tab:mbpp} 
\end{table*}

\noindent{\bf Findings and Implications.} 
The findings from these experiments shed light on the importance of various elements within DocStrings. 
% The removal of newline characters and tabs did not significantly impair the model's efficacy, suggesting that these elements are more stylistic than functional in the context of code generation. 
The removal of newline characters and tabs showed varying impacts across different datasets. While most datasets maintained stable performance, we observed a notable decline in BigCodeBench dataset (from 11.36\% to 7.32\%). This degradation could be attributed to two main reasons: (1) the tokenization mechanism where models may process \code{"""$\backslash$n} differently from \code{"""} alone, potentially affecting the model's ability to properly parse the DocString; and (2) the inherent complexity of BigCodeBench dataset, which presents significant challenges for LLMs, making them more sensitive to any modifications in the input format.

The impact of stop words was also nuanced, with some instances showing a slight Pass@1 change, indicating that even seemingly inconsequential words might play a role in the model's comprehension process. 
The removal of redundant instructions led to a modest improvement in Pass@1 score, supporting our hypothesis that such phrases do not contribute positively to the model's understanding and may even introduce noise.

These experiments underscore the importance of a balanced approach to DocString style. 
While it is beneficial to maintain clear and concise documentation, the removal of certain elements should be done with consideration of their potential impact on the model's comprehension. 

\subsection{Hyper-Parameter Analysis}

\begin{table*}[t]
\centering
\caption{Pass@1 Comparison of Different $\tau$ in CodeHarmony}
\resizebox{0.92\textwidth}{!}{
\begin{tabular}{c|c|c|cccccc}
    \toprule
    \multirow{2}{*}{\textbf{DataSet}} & \multirow{2}{*}{\textbf{$\tau$}} & \multirow{2}{*}{\textbf{Ratio}} & \multicolumn{6}{c}{\textbf{LLMs}} \\
     & & & \textbf{DS-1.3B} & \textbf{DS-6.7B} & \textbf{CQ-7.3B} & \textbf{CG-9.4B} & \textbf{LA-8.0B} & \textbf{Avg.}\\ 
\midrule
\multirow{5}{*}{CodeHarmony} 
 & - & 0 & 59.48 & 64.36 & 60.78 & 64.71 & 59.48 & 61.76\\
 \cmidrule(lr){2-9}
 & 0.985 & 32 & \textbf{58.82} & 62.09 & 59.48 & 56.86 & 56.21 & 58.69 \\
 & 0.990 & 29 & 56.21 & 64.05 & 62.10 & 59.48 & 58.17 & 60.00 \\
 & 0.995 & 26 & 56.86 & \textbf{66.67} & 62.10 & 59.48 & 60.78 & 61.18 \\
 & 0.999 & 25 & 56.21 & 66.01 & \textbf{63.40} & \textbf{60.78} & \textbf{61.44} & \textbf{61.44}\\
\bottomrule
\end{tabular}
}
\label{tab:tau} 
\end{table*}

\begin{table*}[t]
\centering
\caption{Pass@1 Comparison of Different Base Models in CodeHarmony}
\resizebox{1\textwidth}{!}{
\begin{tabular}{c|c|ccccccc}
\toprule
\multirow{2}{*}{\textbf{DataSet}} & \multirow{2}{*}{\textbf{Base Model}} & \multicolumn{6}{c}{\textbf{LLMs}} \\
& & \textbf{DS-1.3B} & \textbf{DS-6.7B} & \textbf{CQ-7.3B} & \textbf{CG-9.4B} & \textbf{LA-8.0B} & \textbf{Avg.}\\ 
\midrule
\multirow{3}{*}{CodeHarmony} 
 & CodeGen-350M & 52.94 & 62.75 & 57.52 & 54.25 & 52.94 & 56.08 \\
 & GPT2 & 41.83 & 45.75 & 47.71 & 43.14 & 44.44 & 44.57 \\
 & CodeGPT-py-adapted & \textbf{56.21} & \textbf{66.01} & \textbf{63.40} & \textbf{60.78} & \textbf{61.44} & \textbf{61.44}\\
\bottomrule
\end{tabular}
}
\label{tab:basemodel} 
\end{table*}

(1) The hyperparameter $\tau$ significantly influences the performance of our method. 
Since the hyperparameter $\tau$ is used to determine the threshold for token importance, it directly affects the number of tokens retained in the compressed DocString, i.e., the compression ratio.
As demonstrated in Table \ref{tab:tau}, different values of $\tau$ result in varying performance metrics across the CodeHarmony dataset. 
The optimal value of $\tau$ that maximizes performance may differ depending on the specific dataset and model architecture. 
For instance, a higher $\tau$ value such as 0.999 seems to work better for the CodeQwen, CodeGeeX4 and Llama3.1 model, while a lower value like 0.985 could be more effective for the DeepSeekCoder-1.3B model. 
Given the good balance or optimal Pass@1 score for the models and datasets, we set the hyperparameter $\tau$ to 0.999.

A limitation in our hyperparameter analysis is that, while we conducted detailed $\tau$ experiments on CodeHarmony (as shown in Table \ref{tab:tau}), we did not perform exhaustive $\tau$ tuning across all datasets.
The optimal $\tau$ value might vary across datasets due to their respective characteristics, and our choice of $\tau=0.999$ based on CodeHarmony may not be optimal for all datasets.
Nevertheless, the experimental results demonstrate that $\tau=0.999$ consistently provides competitive Pass@1 scores across different datasets, as shown in Table~\ref{tab:RQ1}. For instance, {\tool} achieves comparable or better Pass@1 than baseline methods in HumanEval (78.66\%), CodeHarmony (63.40\%) and MBPP (55.40\%) when using the CodeQwen model.

We leave as future work a more comprehensive analysis of $\tau$ across different datasets to better understand its impact on compression performance and to develop dataset-specific tuning strategies.

(2) The choice of base model plays a critical role in the performance of our method. We find that different base models can lead to varying token importance rankings, indicating that the token sorting decisions are not entirely consistent across models. 
This observation raises interesting questions about the relationship between model characteristics and their ability to identify important tokens in DocStrings.

As shown in Table \ref{tab:basemodel}, different base models demonstrate varying baseline performances which can be attributed to their pre-training strategies: CodeGen-350M pre-trained specifically on code corpora outperforms GPT2 pre-trained solely on natural language corpora. 
Notably, CodeGPT-py-adapted achieves the best performance as it combines both approaches; it is built upon GPT2 and further pre-trained on code corpora. This suggests that the ability to understand both natural language and code context is crucial for effective DocString compression.

For the CodeHarmony dataset, while the CodeGPT-py-adapted base model outperforms the other models across all metrics, the token importance rankings produced by different models show notable variations. 
Despite these differences in token sorting, our method maintains robust performance across different base models, suggesting that the overall compression framework is resilient to variations in token importance assessment.

\yg{In summary, we advise users to set 0.999 as the default value of $\tau$ for balanced performance. If users wish to experiment with different compression ratios, they may gradually decrease this value.
Meanwhile, we recommend using the CodeGPT-py-adapted model as the base model for DocString compression, as it consistently achieves the best performance across different datasets.}

% \yg{An intriguing open question remains regarding the minimum requirements for a model to effectively sort tokens: what we might call the "bottom line" for token importance assessment. 
% This includes considerations such as model scale, training data characteristics, and architectural choices that enable effective token sorting while maintaining downstream task performance. 
% We plan to investigate these aspects in future work, aiming to better understand the relationship between model capabilities and effective DocString compression.}

\subsection{Threats to Validity}

In this subsection, we analyze potential threats to the validity of our empirical study.

\noindent\textbf{Threats to Internal Validity.}
The first internal threat is the possibility of implementation errors in {\tool}. To alleviate this, we conducted a thorough code inspection of the implementation and utilized mature libraries. The second internal threat is the implementation correctness of the considered baselines. To mitigate this threat, we implemented all baselines by running their open-source code directly or reimplementing them according to the original studies.

\noindent\textbf{Threats to External Validity.}
The main external threat lies in the choice of datasets and models used in our study. 
For Python code generation, we selected six diverse datasets with high reputations to reflect the complexity of real-world scenarios. However, for cross-language evaluation, we relied on HumanEval-X, which consists of translations of the original HumanEval dataset. The reliance on translated versions of a single dataset, rather than naturally occurring code in different languages, might limit the generalizability of our findings. 
While this choice was constrained by the current availability of multilingual code generation benchmarks, future work could benefit from evaluating on native, language-specific datasets as they become available.

In terms of the choice of models, we prioritized open-source models for better reproducibility and community verification, selecting six representative models that span different scales (from 1.3B to 9.4B parameters) and architectures. These include specialized code models (DeepSeekCoder-1.3b, DeepSeekCoder-6.7b, CodeQwen1.5, CodeGeeX4) and general-purpose models (Llama3.1). 
While proprietary models such as GPT-4 might offer superior performance, their high API costs and frequent version updates could impact the stability and reproducibility of our results. 
Given the hardware constraint (single 3090 GPU) and the need for experimental consistency, we believe our current model selection provides meaningful insights into the effectiveness of our compression method across different scenarios.
% In terms of the choice of models, we selected six state-of-the-art models in code generation including DeepSeekCoder-1.3b, DeepSeekCoder-6.7b, CodeQwen1.5, CodeGeeX4, Llama3.1, and GPT-4o. 

\noindent\textbf{Threats to Construct Validity.}
Construct threats concern the performance metrics used to evaluate {\tool} and baselines. To evaluate the performance of models, we utilized Pass@1 metric which is commonly used in in the previous studies of code generation. Additionally, we used Compression Ratio,  which is also widely used in similar studies, to measure the proportion of reduction in the length of the prompt after compression. 

\yg{Furthermore, we conducted a human study to analyze the qualitative impact of DocString compression, where the Informativeness and Comprehensibility metrics are used which might not capture all aspects of the DocString quality. 
%To mitigate this threat, 
We recruited evaluators with strong Python programming backgrounds and provided them with clear evaluation criteria. 
Additionally, we implemented measures to ensure evaluation quality, including anonymized compressed DocStrings, allowing evaluators to research unfamiliar concepts, and limiting their work load.} 
%number of samples evaluated per session to prevent fatigue.}

%% file: sections/7relate.tex
\section{Related Work}
\label{sec:related}

\subsection{Code Generation} 
Code generation aims to produce code snippets from given natural language descriptions or requirements. Some studies~\cite{mou2015end,ling2016latent} use sequence-based models, which treat the source code as a sequence of tokens and utilize neural networks to generate the source code token by token based on the input description. Other studies~\cite{sun2020treegen,yin2017syntactic} use tree-based models, .ie., construct a parse tree of the program from the natural language description and subsequently convert it into corresponding code.

In recent years, researchers have gradually utilized pre-trained models for code generation tasks, which have outperformed conventional sequence-based and tree-based methods. These models are pre-trained on massive data of source code and then fine-tuned on code generation task. For example, models like CodeGPT~\cite{lu1codexglue}, PLBART~\cite{ahmad2021unified}, and CodeT5~\cite{wang2021codet5} leverage the GPT, BART, and T5 architectures of language models pre-trained on code corpora. However, these models are more suitable for fine-tuning code generation tasks, as their parameter numbers are not large enough to demonstrate emergent capabilities in zero-shot scenarios.

With the development of LLM research, LLMs with over a billion parameters have been employed for zero-shot code generation tasks. Current Code LLMs can be divided into two categories: standard language models and instruction-tuned models~\cite{li2024acecoder}. 

Standard Language models are pre-trained on the raw corpus with the next-token prediction. With the success of GPT series~\cite{brown2020language,radford2018improving,radford2019language} in NLP, Chen et al.~\cite{chen2021evaluating} adapted similar ideas into the domain of source code and fine-tunes GPT models on code to produce closed-source CodeX, which is pre-trained on GitHub code with 12 billion model parameters. To replicate its success, Nijkamp et al.~\cite{nijkamp2022codegen,nijkamp2022codegen} proposed CodeGen and CodeGen2, which are large language models for code with multi-turn program synthesis. Zheng et al.~\cite{zheng2023codegeex} proposed CodeGeeX, a large-scale multilingual code generation model with 13 billion parameters. CodeGeex is pre-trained on a large code corpus of over 20 programming languages and has good performance for generating executable programs in several mainstream programming languages like Python, C++, Java, JavaScript, Go, etc. Li et al.~\cite{li2023starcoder} proposed StarCoder, a 15.5 billion parameter model whose training data incorporates more than 80 different programming languages as well as text extracted from GitHub issues and commits and from notebooks. Differing from the aforementioned decoder-only model, Wang et al.~\cite{wang2023codet5+} proposed CodeT5+, a family of encoder-decoder LLMs for code in which component modules can be flexibly combined to suit a wide range of downstream code tasks. 

Instruction-tuned models are fine-tuned using instruction tuning~\cite{wei2021finetuned}. Instruction tuning helps models follow user's instructions. OpenAI's ChatGPT~\cite{ChatGPT} is trained by Reinforcement Learning with Human Feedback (RLHF)~\cite{ouyang2022training}, making it capable of programming tasks. However, it is of closed-source. For the open-source models, luo et al.~\cite{luo2023wizardcoder} introduce Wizardcoder by fine-tuning StarCoder~\cite{li2023starcoder} with Evol-Instruct and ChatGPT feedback with Code Alpaca’s dataset as seed dataset. wang et al. introduce InstructCodeT5+~\cite{wang2023codet5+} by fine-tuning CodeT5+~\cite{wang2023codet5+} on Code Alpaca’s~\cite{chaudhary2023code} dataset.

In this article, our main goal is to compress DocStrings within code prompts without losing their semantic integrity in the field of code generation. We have observed that the model can still understand the task requirements and generate correct code after removing some of the redundant information in DocStrings. This observation has motivated us to further investigate the compression of DocStrings , an essential component of prompt. To achieve the goal, we have conducted an empirical study on code generation task. 

\subsection{Prompt Compression} 

Prompt Compression attempts to shorten the original prompts without losing essential information. Prompt Compression methods~\cite{jiang2023longllmlingua} can be grouped into three main categories : Token pruning and token merging, Soft prompt tuning methods and Information-entropy-based approaches. Token pruning and token merging need model fine-tuning or intermediate results during inference and have been used
with BERT-scale models. For instance, Modarressi et al.\cite{modarressi2022adapler} introduced AdapLeR which dynamically eliminates less contributing tokens through layers to achieve shorter lengths. Soft prompt tuning requires LLMs' parameter fine-tuning to make them suitable for specific domains but not applicable to black-box LLMs. Wingate et al.~\cite{wingate2022prompt} used the framework of soft prompts to manipulate prompt compression. Mu et al.~\cite{mu2024learning} proposed GIST, a manner to compresse arbitrary prompts into a smaller set of Transformer activations on top of virtual “gist” tokens. 
Kobayashi et al.~\cite{kobayashi2024you} proposed the YORO paradigm, which aims to internalize database knowledge into model parameters to significantly reduce input length and simplify text-to-SQL input.
However, these methods are task-aware and usually tailored for specific tasks and compression ratios, which may limit their generalizability in real-world scenarios.
 
Information-entropy-based approaches use a small language model to calculate the self-information or perplexity of each token in the original prompt and then remove tokens with lower perplexities. For example, Li et al.~\cite{li2023compressing} introduced Selective Context, which employs self-information to filter out less informative content, resulting in the efficiency of the fixed context length. Jiang et al.~\cite{jiang2023llmlingua} proposed LLMLingua, a coarse-to-fine prompt compression method to reduce the length of original prompts while preserving essential information. Based on LLMLingua, Pan et al.\cite{pan-etal-2024-llmlingua} proposed a data distillation procedure to derive knowledge from an LLM (GPT-4) to compress the prompts without losing crucial information. 
All of these approaches are task-agnostic prompt compression methods and have better generalizability and efficiency compared with task-aware methods.

In contrast to the previous studies, we present a novel adaptive compression approach targeting code generation. This compression approach emphasizes the understanding of code semantics and removes redundant content by analyzing the importance of individual tokens. To improve the efficiency of prompt compression, we employ a Top-N strategy, which optimize the compression process and greatly preserve the semantic integrity.

%% file: sections/8conclusion.tex
\section{Conclusion and Future Work}
\label{sec:conclusion}
In our study, we focus on DocString compression and avoiding the loss of essential information in DocString. 
Thereby, we propose a novel compression method {\tool}. This compression method  dynamically adjusts the compression rate and retains greater informativeness and comprehensibility in compressed DocStrings. 
By implementing this compression method, our goal is to  improve model efficiency and reduce the computational resources cost.

While our current research has focused primarily on function level code generation, we recognize that the complexity of code generation increases as the code structure expands. 
Therefore, future work will extend to class level code generation, which will involve more complex logic and larger code structures. 
We believe that {\tool}'s approach can accommodate these higher-level code structures while maintaining its compression efficiency and output quality.

In addition, we plan to explore code generation in a multi-language environment. 
With the diversification of global software development, DocString compression tools that support multiple programming languages will be of higher utility. 
We plan to train a multilingual base language model, which will enhance the applicability and performance of {\tool} in different programming languages.

Furthermore, we aim to extend our research to general code completion tasks. Given the higher frequency of code completion in real-world software engineering compared to natural language to code generation, the motivation for prompt compression becomes even stronger in this context. This extension could potentially bring more significant practical benefits in terms of both computational and economic efficiency.

%% file: main.bbl

\begin{thebibliography}{59}

%%% the Web address.

\ifx \showCODEN    \undefined \def \showCODEN     #1{\unskip}     \fi
\ifx \showDOI      \undefined \def \showDOI       #1{#1}\fi
\ifx \showISBNx    \undefined \def \showISBNx     #1{\unskip}     \fi
\ifx \showISBNxiii \undefined \def \showISBNxiii  #1{\unskip}     \fi
\ifx \showISSN     \undefined \def \showISSN      #1{\unskip}     \fi
\ifx \showLCCN     \undefined \def \showLCCN      #1{\unskip}     \fi
\ifx \shownote     \undefined \def \shownote      #1{#1}          \fi
\ifx \showarticletitle \undefined \def \showarticletitle #1{#1}   \fi
\ifx \showURL      \undefined \def \showURL       {\relax}        \fi
% invisible to TeX
\providecommand\bibfield[2]{#2}
\providecommand\bibinfo[2]{#2}
\providecommand\natexlab[1]{#1}
\providecommand\showeprint[2][]{arXiv:#2}

\bibitem[\protect\citeauthoryear{Ahmad, Chakraborty, Ray, and Chang}{Ahmad et~al\mbox{.}}{2021}]%
        {ahmad2021unified}
\bibfield{author}{\bibinfo{person}{Wasi~Uddin Ahmad}, \bibinfo{person}{Saikat Chakraborty}, \bibinfo{person}{Baishakhi Ray}, {and} \bibinfo{person}{Kai-Wei Chang}.} \bibinfo{year}{2021}\natexlab{}.
\newblock \showarticletitle{Unified pre-training for program understanding and generation}.
\newblock \bibinfo{journal}{\emph{arXiv preprint arXiv:2103.06333}} (\bibinfo{year}{2021}).
\newblock


\bibitem[\protect\citeauthoryear{Austin, Odena, Nye, Bosma, Michalewski, Dohan, Jiang, Cai, Terry, Le, et~al\mbox{.}}{Austin et~al\mbox{.}}{2021}]%
        {austin2021program}
\bibfield{author}{\bibinfo{person}{Jacob Austin}, \bibinfo{person}{Augustus Odena}, \bibinfo{person}{Maxwell Nye}, \bibinfo{person}{Maarten Bosma}, \bibinfo{person}{Henryk Michalewski}, \bibinfo{person}{David Dohan}, \bibinfo{person}{Ellen Jiang}, \bibinfo{person}{Carrie Cai}, \bibinfo{person}{Michael Terry}, \bibinfo{person}{Quoc Le}, {et~al\mbox{.}}} \bibinfo{year}{2021}\natexlab{}.
\newblock \showarticletitle{Program synthesis with large language models}.
\newblock \bibinfo{journal}{\emph{arXiv preprint arXiv:2108.07732}} (\bibinfo{year}{2021}).
\newblock


\bibitem[\protect\citeauthoryear{Bai, Bai, Chu, Cui, Dang, Deng, Fan, Ge, Han, Huang, et~al\mbox{.}}{Bai et~al\mbox{.}}{2023}]%
        {bai2023qwen}
\bibfield{author}{\bibinfo{person}{Jinze Bai}, \bibinfo{person}{Shuai Bai}, \bibinfo{person}{Yunfei Chu}, \bibinfo{person}{Zeyu Cui}, \bibinfo{person}{Kai Dang}, \bibinfo{person}{Xiaodong Deng}, \bibinfo{person}{Yang Fan}, \bibinfo{person}{Wenbin Ge}, \bibinfo{person}{Yu Han}, \bibinfo{person}{Fei Huang}, {et~al\mbox{.}}} \bibinfo{year}{2023}\natexlab{}.
\newblock \showarticletitle{Qwen technical report}.
\newblock \bibinfo{journal}{\emph{arXiv preprint arXiv:2309.16609}} (\bibinfo{year}{2023}).
\newblock


\bibitem[\protect\citeauthoryear{Brown}{Brown}{2020}]%
        {brown2020language}
\bibfield{author}{\bibinfo{person}{Tom~B Brown}.} \bibinfo{year}{2020}\natexlab{}.
\newblock \showarticletitle{Language models are few-shot learners}.
\newblock \bibinfo{journal}{\emph{arXiv preprint arXiv:2005.14165}} (\bibinfo{year}{2020}).
\newblock


\bibitem[\protect\citeauthoryear{Chaudhary}{Chaudhary}{2023}]%
        {chaudhary2023code}
\bibfield{author}{\bibinfo{person}{Sahil Chaudhary}.} \bibinfo{year}{2023}\natexlab{}.
\newblock \showarticletitle{Code alpaca: An instruction-following llama model for code generation}.
\newblock \bibinfo{journal}{\emph{GitHub repository}} (\bibinfo{year}{2023}).
\newblock


\bibitem[\protect\citeauthoryear{Chen, Tworek, Jun, Yuan, Pinto, Kaplan, Edwards, Burda, Joseph, Brockman, et~al\mbox{.}}{Chen et~al\mbox{.}}{2021}]%
        {chen2021evaluating}
\bibfield{author}{\bibinfo{person}{Mark Chen}, \bibinfo{person}{Jerry Tworek}, \bibinfo{person}{Heewoo Jun}, \bibinfo{person}{Qiming Yuan}, \bibinfo{person}{Henrique Ponde De~Oliveira Pinto}, \bibinfo{person}{Jared Kaplan}, \bibinfo{person}{Harri Edwards}, \bibinfo{person}{Yuri Burda}, \bibinfo{person}{Nicholas Joseph}, \bibinfo{person}{Greg Brockman}, {et~al\mbox{.}}} \bibinfo{year}{2021}\natexlab{}.
\newblock \showarticletitle{Evaluating large language models trained on code}.
\newblock \bibinfo{journal}{\emph{arXiv preprint arXiv:2107.03374}} (\bibinfo{year}{2021}).
\newblock


\bibitem[\protect\citeauthoryear{Chen, Hu, Zhi, Han, Deng, and Yin}{Chen et~al\mbox{.}}{2024}]%
        {chen2024chatunitest}
\bibfield{author}{\bibinfo{person}{Yinghao Chen}, \bibinfo{person}{Zehao Hu}, \bibinfo{person}{Chen Zhi}, \bibinfo{person}{Junxiao Han}, \bibinfo{person}{Shuiguang Deng}, {and} \bibinfo{person}{Jianwei Yin}.} \bibinfo{year}{2024}\natexlab{}.
\newblock \showarticletitle{Chatunitest: A framework for llm-based test generation}. In \bibinfo{booktitle}{\emph{Companion Proceedings of the 32nd ACM International Conference on the Foundations of Software Engineering}}. \bibinfo{pages}{572--576}.
\newblock


\bibitem[\protect\citeauthoryear{Dainese, Ilin, and Marttinen}{Dainese et~al\mbox{.}}{2024}]%
        {dainese2024can}
\bibfield{author}{\bibinfo{person}{Nicola Dainese}, \bibinfo{person}{Alexander Ilin}, {and} \bibinfo{person}{Pekka Marttinen}.} \bibinfo{year}{2024}\natexlab{}.
\newblock \showarticletitle{Can docstring reformulation with an LLM improve code generation?}. In \bibinfo{booktitle}{\emph{Proceedings of the 18th Conference of the European Chapter of the Association for Computational Linguistics: Student Research Workshop}}. \bibinfo{pages}{296--312}.
\newblock


\bibitem[\protect\citeauthoryear{Ding, Peng, Chen, Huang, Bian, and Zheng}{Ding et~al\mbox{.}}{2024}]%
        {ding2024code}
\bibfield{author}{\bibinfo{person}{Xi Ding}, \bibinfo{person}{Rui Peng}, \bibinfo{person}{Xiangping Chen}, \bibinfo{person}{Yuan Huang}, \bibinfo{person}{Jing Bian}, {and} \bibinfo{person}{Zibin Zheng}.} \bibinfo{year}{2024}\natexlab{}.
\newblock \showarticletitle{Do Code Summarization Models Process Too Much Information? Function Signature May Be All That Is Needed}.
\newblock \bibinfo{journal}{\emph{ACM Transactions on Software Engineering and Methodology}} \bibinfo{volume}{33}, \bibinfo{number}{6} (\bibinfo{year}{2024}), \bibinfo{pages}{1--35}.
\newblock


\bibitem[\protect\citeauthoryear{Dubey, Jauhri, Pandey, Kadian, Al-Dahle, Letman, Mathur, Schelten, Yang, Fan, et~al\mbox{.}}{Dubey et~al\mbox{.}}{2024}]%
        {dubey2024llama}
\bibfield{author}{\bibinfo{person}{Abhimanyu Dubey}, \bibinfo{person}{Abhinav Jauhri}, \bibinfo{person}{Abhinav Pandey}, \bibinfo{person}{Abhishek Kadian}, \bibinfo{person}{Ahmad Al-Dahle}, \bibinfo{person}{Aiesha Letman}, \bibinfo{person}{Akhil Mathur}, \bibinfo{person}{Alan Schelten}, \bibinfo{person}{Amy Yang}, \bibinfo{person}{Angela Fan}, {et~al\mbox{.}}} \bibinfo{year}{2024}\natexlab{}.
\newblock \showarticletitle{The llama 3 herd of models}.
\newblock \bibinfo{journal}{\emph{arXiv preprint arXiv:2407.21783}} (\bibinfo{year}{2024}).
\newblock


\bibitem[\protect\citeauthoryear{Dvivedi, Vijay, Pujari, Lodh, and Kumar}{Dvivedi et~al\mbox{.}}{2024}]%
        {dvivedi2024comparative}
\bibfield{author}{\bibinfo{person}{Shubhang~Shekhar Dvivedi}, \bibinfo{person}{Vyshnav Vijay}, \bibinfo{person}{Sai Leela~Rahul Pujari}, \bibinfo{person}{Shoumik Lodh}, {and} \bibinfo{person}{Dhruv Kumar}.} \bibinfo{year}{2024}\natexlab{}.
\newblock \showarticletitle{A comparative analysis of large language models for code documentation generation}. In \bibinfo{booktitle}{\emph{Proceedings of the 1st ACM International Conference on AI-Powered Software}}. \bibinfo{pages}{65--73}.
\newblock


\bibitem[\protect\citeauthoryear{Flab-Pruner}{Flab-Pruner}{2024}]%
        {Flab-Pruner}
\bibfield{author}{\bibinfo{person}{Flab-Pruner}.} \bibinfo{year}{2024}\natexlab{}.
\newblock \bibinfo{title}{Flab-Pruner: Towards Greener Yet Powerful Code Intelligence via Structural Pruning}.
\newblock \bibinfo{howpublished}{\url{https://github.com/Flab-Pruner/Flab-Pruner}}.
\newblock


\bibitem[\protect\citeauthoryear{Fleiss}{Fleiss}{1971}]%
        {fleiss1971measuring}
\bibfield{author}{\bibinfo{person}{Joseph~L Fleiss}.} \bibinfo{year}{1971}\natexlab{}.
\newblock \showarticletitle{Measuring nominal scale agreement among many raters.}
\newblock \bibinfo{journal}{\emph{Psychological bulletin}} \bibinfo{volume}{76}, \bibinfo{number}{5} (\bibinfo{year}{1971}), \bibinfo{pages}{378}.
\newblock


\bibitem[\protect\citeauthoryear{Guo, Zhu, Yang, Xie, Dong, Zhang, Chen, Bi, Li, et~al\mbox{.}}{Guo et~al\mbox{.}}{2024}]%
        {guo2024deepseek}
\bibfield{author}{\bibinfo{person}{Daya Guo}, \bibinfo{person}{Qihao Zhu}, \bibinfo{person}{Dejian Yang}, \bibinfo{person}{Zhenda Xie}, \bibinfo{person}{Kai Dong}, \bibinfo{person}{Wentao Zhang}, \bibinfo{person}{Guanting Chen}, \bibinfo{person}{Xiao Bi}, \bibinfo{person}{YK Li}, {et~al\mbox{.}}} \bibinfo{year}{2024}\natexlab{}.
\newblock \showarticletitle{DeepSeek-Coder: When the Large Language Model Meets Programming--The Rise of Code Intelligence}.
\newblock \bibinfo{journal}{\emph{arXiv preprint arXiv:2401.14196}} (\bibinfo{year}{2024}).
\newblock


\bibitem[\protect\citeauthoryear{Jiang, Wu, Lin, Yang, and Qiu}{Jiang et~al\mbox{.}}{2023a}]%
        {jiang2023llmlingua}
\bibfield{author}{\bibinfo{person}{Huiqiang Jiang}, \bibinfo{person}{Qianhui Wu}, \bibinfo{person}{Chin-Yew Lin}, \bibinfo{person}{Yuqing Yang}, {and} \bibinfo{person}{Lili Qiu}.} \bibinfo{year}{2023}\natexlab{a}.
\newblock \showarticletitle{LLMLingua: Compressing Prompts for Accelerated Inference of Large Language Models}. In \bibinfo{booktitle}{\emph{Proceedings of the 2023 Conference on Empirical Methods in Natural Language Processing}}. \bibinfo{pages}{13358--13376}.
\newblock


\bibitem[\protect\citeauthoryear{Jiang, Wu, Luo, Li, Lin, Yang, and Qiu}{Jiang et~al\mbox{.}}{2023b}]%
        {jiang2023longllmlingua}
\bibfield{author}{\bibinfo{person}{Huiqiang Jiang}, \bibinfo{person}{Qianhui Wu}, \bibinfo{person}{Xufang Luo}, \bibinfo{person}{Dongsheng Li}, \bibinfo{person}{Chin-Yew Lin}, \bibinfo{person}{Yuqing Yang}, {and} \bibinfo{person}{Lili Qiu}.} \bibinfo{year}{2023}\natexlab{b}.
\newblock \showarticletitle{Longllmlingua: Accelerating and enhancing llms in long context scenarios via prompt compression}.
\newblock \bibinfo{journal}{\emph{arXiv preprint arXiv:2310.06839}} (\bibinfo{year}{2023}).
\newblock


\bibitem[\protect\citeauthoryear{Kobayashi, Lan, Shi, Chang, Guo, Zhu, Wang, and Ng}{Kobayashi et~al\mbox{.}}{2024}]%
        {kobayashi2024you}
\bibfield{author}{\bibinfo{person}{Hideo Kobayashi}, \bibinfo{person}{Wuwei Lan}, \bibinfo{person}{Peng Shi}, \bibinfo{person}{Shuaichen Chang}, \bibinfo{person}{Jiang Guo}, \bibinfo{person}{Henghui Zhu}, \bibinfo{person}{Zhiguo Wang}, {and} \bibinfo{person}{Patrick Ng}.} \bibinfo{year}{2024}\natexlab{}.
\newblock \showarticletitle{You Only Read Once (YORO): Learning to Internalize Database Knowledge for Text-to-SQL}.
\newblock \bibinfo{journal}{\emph{arXiv preprint arXiv:2409.12172}} (\bibinfo{year}{2024}).
\newblock


\bibitem[\protect\citeauthoryear{Kwon, Li, Zhuang, Sheng, Zheng, Yu, Gonzalez, Zhang, and Stoica}{Kwon et~al\mbox{.}}{2023}]%
        {kwon2023efficient}
\bibfield{author}{\bibinfo{person}{Woosuk Kwon}, \bibinfo{person}{Zhuohan Li}, \bibinfo{person}{Siyuan Zhuang}, \bibinfo{person}{Ying Sheng}, \bibinfo{person}{Lianmin Zheng}, \bibinfo{person}{Cody~Hao Yu}, \bibinfo{person}{Joseph~E. Gonzalez}, \bibinfo{person}{Hao Zhang}, {and} \bibinfo{person}{Ion Stoica}.} \bibinfo{year}{2023}\natexlab{}.
\newblock \showarticletitle{Efficient Memory Management for Large Language Model Serving with PagedAttention}. In \bibinfo{booktitle}{\emph{Proceedings of the ACM SIGOPS 29th Symposium on Operating Systems Principles}}.
\newblock


\bibitem[\protect\citeauthoryear{Li, Zhao, Li, Li, and Jin}{Li et~al\mbox{.}}{2024}]%
        {li2024acecoder}
\bibfield{author}{\bibinfo{person}{Jia Li}, \bibinfo{person}{Yunfei Zhao}, \bibinfo{person}{Yongmin Li}, \bibinfo{person}{Ge Li}, {and} \bibinfo{person}{Zhi Jin}.} \bibinfo{year}{2024}\natexlab{}.
\newblock \showarticletitle{AceCoder: An Effective Prompting Technique Specialized in Code Generation}.
\newblock \bibinfo{journal}{\emph{ACM Transactions on Software Engineering and Methodology}} (\bibinfo{year}{2024}).
\newblock


\bibitem[\protect\citeauthoryear{Li, Allal, Zi, Muennighoff, Kocetkov, Mou, Marone, Akiki, Li, Chim, et~al\mbox{.}}{Li et~al\mbox{.}}{2023a}]%
        {li2023starcoder}
\bibfield{author}{\bibinfo{person}{Raymond Li}, \bibinfo{person}{Loubna~Ben Allal}, \bibinfo{person}{Yangtian Zi}, \bibinfo{person}{Niklas Muennighoff}, \bibinfo{person}{Denis Kocetkov}, \bibinfo{person}{Chenghao Mou}, \bibinfo{person}{Marc Marone}, \bibinfo{person}{Christopher Akiki}, \bibinfo{person}{Jia Li}, \bibinfo{person}{Jenny Chim}, {et~al\mbox{.}}} \bibinfo{year}{2023}\natexlab{a}.
\newblock \showarticletitle{Starcoder: may the source be with you!}
\newblock \bibinfo{journal}{\emph{arXiv preprint arXiv:2305.06161}} (\bibinfo{year}{2023}).
\newblock


\bibitem[\protect\citeauthoryear{Li, Choi, Chung, Kushman, Schrittwieser, Leblond, Eccles, Keeling, Gimeno, Dal~Lago, et~al\mbox{.}}{Li et~al\mbox{.}}{2022}]%
        {li2022competition}
\bibfield{author}{\bibinfo{person}{Yujia Li}, \bibinfo{person}{David Choi}, \bibinfo{person}{Junyoung Chung}, \bibinfo{person}{Nate Kushman}, \bibinfo{person}{Julian Schrittwieser}, \bibinfo{person}{R{\'e}mi Leblond}, \bibinfo{person}{Tom Eccles}, \bibinfo{person}{James Keeling}, \bibinfo{person}{Felix Gimeno}, \bibinfo{person}{Agustin Dal~Lago}, {et~al\mbox{.}}} \bibinfo{year}{2022}\natexlab{}.
\newblock \showarticletitle{Competition-level code generation with alphacode}.
\newblock \bibinfo{journal}{\emph{Science}} \bibinfo{volume}{378}, \bibinfo{number}{6624} (\bibinfo{year}{2022}), \bibinfo{pages}{1092--1097}.
\newblock


\bibitem[\protect\citeauthoryear{Li, Dong, Guerin, and Lin}{Li et~al\mbox{.}}{2023b}]%
        {li2023compressing}
\bibfield{author}{\bibinfo{person}{Yucheng Li}, \bibinfo{person}{Bo Dong}, \bibinfo{person}{Frank Guerin}, {and} \bibinfo{person}{Chenghua Lin}.} \bibinfo{year}{2023}\natexlab{b}.
\newblock \showarticletitle{Compressing Context to Enhance Inference Efficiency of Large Language Models}. In \bibinfo{booktitle}{\emph{Proceedings of the 2023 Conference on Empirical Methods in Natural Language Processing}}. \bibinfo{pages}{6342--6353}.
\newblock


\bibitem[\protect\citeauthoryear{Ling, Grefenstette, Hermann, Ko{\v{c}}isk{\`y}, Senior, Wang, and Blunsom}{Ling et~al\mbox{.}}{2016}]%
        {ling2016latent}
\bibfield{author}{\bibinfo{person}{Wang Ling}, \bibinfo{person}{Edward Grefenstette}, \bibinfo{person}{Karl~Moritz Hermann}, \bibinfo{person}{Tom{\'a}{\v{s}} Ko{\v{c}}isk{\`y}}, \bibinfo{person}{Andrew Senior}, \bibinfo{person}{Fumin Wang}, {and} \bibinfo{person}{Phil Blunsom}.} \bibinfo{year}{2016}\natexlab{}.
\newblock \showarticletitle{Latent predictor networks for code generation}.
\newblock \bibinfo{journal}{\emph{arXiv preprint arXiv:1603.06744}} (\bibinfo{year}{2016}).
\newblock


\bibitem[\protect\citeauthoryear{Liu, Xia, Wang, and Zhang}{Liu et~al\mbox{.}}{2024}]%
        {liu2024your}
\bibfield{author}{\bibinfo{person}{Jiawei Liu}, \bibinfo{person}{Chunqiu~Steven Xia}, \bibinfo{person}{Yuyao Wang}, {and} \bibinfo{person}{Lingming Zhang}.} \bibinfo{year}{2024}\natexlab{}.
\newblock \showarticletitle{Is your code generated by chatgpt really correct? rigorous evaluation of large language models for code generation}.
\newblock \bibinfo{journal}{\emph{Advances in Neural Information Processing Systems}}  \bibinfo{volume}{36} (\bibinfo{year}{2024}).
\newblock


\bibitem[\protect\citeauthoryear{Lu, Guo, Ren, Huang, Svyatkovskiy, Blanco, Clement, Drain, Jiang, Tang, et~al\mbox{.}}{Lu et~al\mbox{.}}{[n.d.]}]%
        {lu1codexglue}
\bibfield{author}{\bibinfo{person}{Shuai Lu}, \bibinfo{person}{Daya Guo}, \bibinfo{person}{Shuo Ren}, \bibinfo{person}{Junjie Huang}, \bibinfo{person}{Alexey Svyatkovskiy}, \bibinfo{person}{Ambrosio Blanco}, \bibinfo{person}{Colin Clement}, \bibinfo{person}{Dawn Drain}, \bibinfo{person}{Daxin Jiang}, \bibinfo{person}{Duyu Tang}, {et~al\mbox{.}}} \bibinfo{year}{[n.d.]}\natexlab{}.
\newblock \showarticletitle{CodeXGLUE: A Machine Learning Benchmark Dataset for Code Understanding and Generation}. In \bibinfo{booktitle}{\emph{Thirty-fifth Conference on Neural Information Processing Systems Datasets and Benchmarks Track (Round 1)}}.
\newblock


\bibitem[\protect\citeauthoryear{Luo, Xu, Zhao, Sun, Geng, Hu, Tao, Ma, Lin, and Jiang}{Luo et~al\mbox{.}}{2023}]%
        {luo2023wizardcoder}
\bibfield{author}{\bibinfo{person}{Ziyang Luo}, \bibinfo{person}{Can Xu}, \bibinfo{person}{Pu Zhao}, \bibinfo{person}{Qingfeng Sun}, \bibinfo{person}{Xiubo Geng}, \bibinfo{person}{Wenxiang Hu}, \bibinfo{person}{Chongyang Tao}, \bibinfo{person}{Jing Ma}, \bibinfo{person}{Qingwei Lin}, {and} \bibinfo{person}{Daxin Jiang}.} \bibinfo{year}{2023}\natexlab{}.
\newblock \showarticletitle{Wizardcoder: Empowering code large language models with evol-instruct}.
\newblock \bibinfo{journal}{\emph{arXiv preprint arXiv:2306.08568}} (\bibinfo{year}{2023}).
\newblock


\bibitem[\protect\citeauthoryear{Miceli-Barone and Sennrich}{Miceli-Barone and Sennrich}{2017}]%
        {miceli2017parallel}
\bibfield{author}{\bibinfo{person}{Antonio~Valerio Miceli-Barone} {and} \bibinfo{person}{Rico Sennrich}.} \bibinfo{year}{2017}\natexlab{}.
\newblock \showarticletitle{A Parallel Corpus of Python Functions and Documentation Strings for Automated Code Documentation and Code Generation}. In \bibinfo{booktitle}{\emph{Proceedings of the Eighth International Joint Conference on Natural Language Processing (Volume 2: Short Papers)}}. \bibinfo{pages}{314--319}.
\newblock


\bibitem[\protect\citeauthoryear{Modarressi, Mohebbi, and Pilehvar}{Modarressi et~al\mbox{.}}{2022}]%
        {modarressi2022adapler}
\bibfield{author}{\bibinfo{person}{Ali Modarressi}, \bibinfo{person}{Hosein Mohebbi}, {and} \bibinfo{person}{Mohammad~Taher Pilehvar}.} \bibinfo{year}{2022}\natexlab{}.
\newblock \showarticletitle{Adapler: Speeding up inference by adaptive length reduction}.
\newblock \bibinfo{journal}{\emph{arXiv preprint arXiv:2203.08991}} (\bibinfo{year}{2022}).
\newblock


\bibitem[\protect\citeauthoryear{Mou, Men, Li, Zhang, and Jin}{Mou et~al\mbox{.}}{2015}]%
        {mou2015end}
\bibfield{author}{\bibinfo{person}{Lili Mou}, \bibinfo{person}{Rui Men}, \bibinfo{person}{Ge Li}, \bibinfo{person}{Lu Zhang}, {and} \bibinfo{person}{Zhi Jin}.} \bibinfo{year}{2015}\natexlab{}.
\newblock \showarticletitle{On end-to-end program generation from user intention by deep neural networks}.
\newblock \bibinfo{journal}{\emph{arXiv preprint arXiv:1510.07211}} (\bibinfo{year}{2015}).
\newblock


\bibitem[\protect\citeauthoryear{Mu, Li, and Goodman}{Mu et~al\mbox{.}}{2024}]%
        {mu2024learning}
\bibfield{author}{\bibinfo{person}{Jesse Mu}, \bibinfo{person}{Xiang Li}, {and} \bibinfo{person}{Noah Goodman}.} \bibinfo{year}{2024}\natexlab{}.
\newblock \showarticletitle{Learning to compress prompts with gist tokens}.
\newblock \bibinfo{journal}{\emph{Advances in Neural Information Processing Systems}}  \bibinfo{volume}{36} (\bibinfo{year}{2024}).
\newblock


\bibitem[\protect\citeauthoryear{Nam, Macvean, Hellendoorn, Vasilescu, and Myers}{Nam et~al\mbox{.}}{2024}]%
        {nam2024using}
\bibfield{author}{\bibinfo{person}{Daye Nam}, \bibinfo{person}{Andrew Macvean}, \bibinfo{person}{Vincent Hellendoorn}, \bibinfo{person}{Bogdan Vasilescu}, {and} \bibinfo{person}{Brad Myers}.} \bibinfo{year}{2024}\natexlab{}.
\newblock \showarticletitle{Using an llm to help with code understanding}. In \bibinfo{booktitle}{\emph{Proceedings of the IEEE/ACM 46th International Conference on Software Engineering}}. \bibinfo{pages}{1--13}.
\newblock


\bibitem[\protect\citeauthoryear{Neelakantan, Xu, Puri, Radford, Han, Tworek, Yuan, Tezak, Kim, Hallacy, et~al\mbox{.}}{Neelakantan et~al\mbox{.}}{2022}]%
        {neelakantan2022text}
\bibfield{author}{\bibinfo{person}{Arvind Neelakantan}, \bibinfo{person}{Tao Xu}, \bibinfo{person}{Raul Puri}, \bibinfo{person}{Alec Radford}, \bibinfo{person}{Jesse~Michael Han}, \bibinfo{person}{Jerry Tworek}, \bibinfo{person}{Qiming Yuan}, \bibinfo{person}{Nikolas Tezak}, \bibinfo{person}{Jong~Wook Kim}, \bibinfo{person}{Chris Hallacy}, {et~al\mbox{.}}} \bibinfo{year}{2022}\natexlab{}.
\newblock \showarticletitle{Text and code embeddings by contrastive pre-training}.
\newblock \bibinfo{journal}{\emph{arXiv preprint arXiv:2201.10005}} (\bibinfo{year}{2022}).
\newblock


\bibitem[\protect\citeauthoryear{Nijkamp, Pang, Hayashi, Tu, Wang, Zhou, Savarese, and Xiong}{Nijkamp et~al\mbox{.}}{2022}]%
        {nijkamp2022codegen}
\bibfield{author}{\bibinfo{person}{Erik Nijkamp}, \bibinfo{person}{Bo Pang}, \bibinfo{person}{Hiroaki Hayashi}, \bibinfo{person}{Lifu Tu}, \bibinfo{person}{Huan Wang}, \bibinfo{person}{Yingbo Zhou}, \bibinfo{person}{Silvio Savarese}, {and} \bibinfo{person}{Caiming Xiong}.} \bibinfo{year}{2022}\natexlab{}.
\newblock \showarticletitle{Codegen: An open large language model for code with multi-turn program synthesis}.
\newblock \bibinfo{journal}{\emph{arXiv preprint arXiv:2203.13474}} (\bibinfo{year}{2022}).
\newblock


\bibitem[\protect\citeauthoryear{Niu, Zhang, Li, Luo, and Ng}{Niu et~al\mbox{.}}{2024}]%
        {niu2024evaluating}
\bibfield{author}{\bibinfo{person}{Changan Niu}, \bibinfo{person}{Ting Zhang}, \bibinfo{person}{Chuanyi Li}, \bibinfo{person}{Bin Luo}, {and} \bibinfo{person}{Vincent Ng}.} \bibinfo{year}{2024}\natexlab{}.
\newblock \showarticletitle{On Evaluating the Efficiency of Source Code Generated by LLMs}. In \bibinfo{booktitle}{\emph{Proceedings of the 2024 IEEE/ACM First International Conference on AI Foundation Models and Software Engineering}}. \bibinfo{pages}{103--107}.
\newblock


\bibitem[\protect\citeauthoryear{OpenAI}{OpenAI}{2022}]%
        {ChatGPT}
\bibfield{author}{\bibinfo{person}{OpenAI}.} \bibinfo{year}{2022}\natexlab{}.
\newblock \bibinfo{title}{ChatGPT}.
\newblock
\newblock
\newblock
\shownote{\url{https://openai.com/blog/ chatgpt}.}


\bibitem[\protect\citeauthoryear{Ouyang, Wu, Jiang, Almeida, Wainwright, Mishkin, Zhang, Agarwal, Slama, Ray, et~al\mbox{.}}{Ouyang et~al\mbox{.}}{2022}]%
        {ouyang2022training}
\bibfield{author}{\bibinfo{person}{Long Ouyang}, \bibinfo{person}{Jeffrey Wu}, \bibinfo{person}{Xu Jiang}, \bibinfo{person}{Diogo Almeida}, \bibinfo{person}{Carroll Wainwright}, \bibinfo{person}{Pamela Mishkin}, \bibinfo{person}{Chong Zhang}, \bibinfo{person}{Sandhini Agarwal}, \bibinfo{person}{Katarina Slama}, \bibinfo{person}{Alex Ray}, {et~al\mbox{.}}} \bibinfo{year}{2022}\natexlab{}.
\newblock \showarticletitle{Training language models to follow instructions with human feedback}.
\newblock \bibinfo{journal}{\emph{Advances in neural information processing systems}}  \bibinfo{volume}{35} (\bibinfo{year}{2022}), \bibinfo{pages}{27730--27744}.
\newblock


\bibitem[\protect\citeauthoryear{Pan, Wu, Jiang, Xia, Luo, Zhang, Lin, R{\"u}hle, Yang, Lin, Zhao, Qiu, and Zhang}{Pan et~al\mbox{.}}{2024}]%
        {pan-etal-2024-llmlingua}
\bibfield{author}{\bibinfo{person}{Zhuoshi Pan}, \bibinfo{person}{Qianhui Wu}, \bibinfo{person}{Huiqiang Jiang}, \bibinfo{person}{Menglin Xia}, \bibinfo{person}{Xufang Luo}, \bibinfo{person}{Jue Zhang}, \bibinfo{person}{Qingwei Lin}, \bibinfo{person}{Victor R{\"u}hle}, \bibinfo{person}{Yuqing Yang}, \bibinfo{person}{Chin-Yew Lin}, \bibinfo{person}{H.~Vicky Zhao}, \bibinfo{person}{Lili Qiu}, {and} \bibinfo{person}{Dongmei Zhang}.} \bibinfo{year}{2024}\natexlab{}.
\newblock \showarticletitle{{LLML}ingua-2: Data Distillation for Efficient and Faithful Task-Agnostic Prompt Compression}. In \bibinfo{booktitle}{\emph{Findings of the Association for Computational Linguistics ACL 2024}}, \bibfield{editor}{\bibinfo{person}{Lun-Wei Ku}, \bibinfo{person}{Andre Martins}, {and} \bibinfo{person}{Vivek Srikumar}} (Eds.). \bibinfo{publisher}{Association for Computational Linguistics}, \bibinfo{address}{Bangkok, Thailand and virtual meeting}, \bibinfo{pages}{963--981}.
\newblock
\urldef\tempurl%
\url{https://aclanthology.org/2024.findings-acl.57}
\showURL{%
\tempurl}


\bibitem[\protect\citeauthoryear{Poudel, Cook, Traore, and Ameli}{Poudel et~al\mbox{.}}{2024}]%
        {poudel2024documint}
\bibfield{author}{\bibinfo{person}{Bibek Poudel}, \bibinfo{person}{Adam Cook}, \bibinfo{person}{Sekou Traore}, {and} \bibinfo{person}{Shelah Ameli}.} \bibinfo{year}{2024}\natexlab{}.
\newblock \showarticletitle{DocuMint: Docstring Generation for Python using Small Language Models}.
\newblock \bibinfo{journal}{\emph{arXiv preprint arXiv:2405.10243}} (\bibinfo{year}{2024}).
\newblock


\bibitem[\protect\citeauthoryear{Radford}{Radford}{2018}]%
        {radford2018improving}
\bibfield{author}{\bibinfo{person}{Alec Radford}.} \bibinfo{year}{2018}\natexlab{}.
\newblock \showarticletitle{Improving language understanding by generative pre-training}.
\newblock  (\bibinfo{year}{2018}).
\newblock


\bibitem[\protect\citeauthoryear{Radford, Wu, Child, Luan, Amodei, Sutskever, et~al\mbox{.}}{Radford et~al\mbox{.}}{2019}]%
        {radford2019language}
\bibfield{author}{\bibinfo{person}{Alec Radford}, \bibinfo{person}{Jeffrey Wu}, \bibinfo{person}{Rewon Child}, \bibinfo{person}{David Luan}, \bibinfo{person}{Dario Amodei}, \bibinfo{person}{Ilya Sutskever}, {et~al\mbox{.}}} \bibinfo{year}{2019}\natexlab{}.
\newblock \showarticletitle{Language models are unsupervised multitask learners}.
\newblock \bibinfo{journal}{\emph{OpenAI blog}} \bibinfo{volume}{1}, \bibinfo{number}{8} (\bibinfo{year}{2019}), \bibinfo{pages}{9}.
\newblock


\bibitem[\protect\citeauthoryear{Randolph}{Randolph}{2005}]%
        {randolph2005free}
\bibfield{author}{\bibinfo{person}{Justus~J Randolph}.} \bibinfo{year}{2005}\natexlab{}.
\newblock \showarticletitle{Free-Marginal Multirater Kappa (multirater K [free]): An Alternative to Fleiss' Fixed-Marginal Multirater Kappa.}
\newblock \bibinfo{journal}{\emph{Online submission}} (\bibinfo{year}{2005}).
\newblock


\bibitem[\protect\citeauthoryear{Rasnayaka, Wang, Shariffdeen, and Iyer}{Rasnayaka et~al\mbox{.}}{2024}]%
        {rasnayaka2024empirical}
\bibfield{author}{\bibinfo{person}{Sanka Rasnayaka}, \bibinfo{person}{Guanlin Wang}, \bibinfo{person}{Ridwan Shariffdeen}, {and} \bibinfo{person}{Ganesh~Neelakanta Iyer}.} \bibinfo{year}{2024}\natexlab{}.
\newblock \showarticletitle{An empirical study on usage and perceptions of llms in a software engineering project}. In \bibinfo{booktitle}{\emph{Proceedings of the 1st International Workshop on Large Language Models for Code}}. \bibinfo{pages}{111--118}.
\newblock


\bibitem[\protect\citeauthoryear{Shi, Yang, Kang, Xu, He, and Lo}{Shi et~al\mbox{.}}{2024}]%
        {shi2024greening}
\bibfield{author}{\bibinfo{person}{Jieke Shi}, \bibinfo{person}{Zhou Yang}, \bibinfo{person}{Hong~Jin Kang}, \bibinfo{person}{Bowen Xu}, \bibinfo{person}{Junda He}, {and} \bibinfo{person}{David Lo}.} \bibinfo{year}{2024}\natexlab{}.
\newblock \showarticletitle{Greening large language models of code}. In \bibinfo{booktitle}{\emph{Proceedings of the 46th International Conference on Software Engineering: Software Engineering in Society}}. \bibinfo{pages}{142--153}.
\newblock


\bibitem[\protect\citeauthoryear{Sun, Du, Song, Wang, and Li}{Sun et~al\mbox{.}}{2024a}]%
        {sun2024neural}
\bibfield{author}{\bibinfo{person}{Zhensu Sun}, \bibinfo{person}{Xiaoning Du}, \bibinfo{person}{Fu Song}, \bibinfo{person}{Shangwen Wang}, {and} \bibinfo{person}{Li Li}.} \bibinfo{year}{2024}\natexlab{a}.
\newblock \showarticletitle{When Neural Code Completion Models Size up the Situation: Attaining Cheaper and Faster Completion through Dynamic Model Inference}. In \bibinfo{booktitle}{\emph{Proceedings of the IEEE/ACM 46th International Conference on Software Engineering}}. \bibinfo{pages}{1--12}.
\newblock


\bibitem[\protect\citeauthoryear{Sun, Du, Song, Wang, Ni, Li, and Lo}{Sun et~al\mbox{.}}{2024b}]%
        {sun2024don}
\bibfield{author}{\bibinfo{person}{Zhensu Sun}, \bibinfo{person}{Xiaoning Du}, \bibinfo{person}{Fu Song}, \bibinfo{person}{Shangwen Wang}, \bibinfo{person}{Mingze Ni}, \bibinfo{person}{Li Li}, {and} \bibinfo{person}{David Lo}.} \bibinfo{year}{2024}\natexlab{b}.
\newblock \showarticletitle{Don’t Complete It! Preventing Unhelpful Code Completion for Productive and Sustainable Neural Code Completion Systems}.
\newblock \bibinfo{journal}{\emph{ACM Transactions on Software Engineering and Methodology}} \bibinfo{volume}{34}, \bibinfo{number}{1} (\bibinfo{year}{2024}), \bibinfo{pages}{1--22}.
\newblock


\bibitem[\protect\citeauthoryear{Sun, Du, Yang, Li, and Lo}{Sun et~al\mbox{.}}{2024c}]%
        {sun2024ai}
\bibfield{author}{\bibinfo{person}{Zhensu Sun}, \bibinfo{person}{Xiaoning Du}, \bibinfo{person}{Zhou Yang}, \bibinfo{person}{Li Li}, {and} \bibinfo{person}{David Lo}.} \bibinfo{year}{2024}\natexlab{c}.
\newblock \showarticletitle{Ai coders are among us: Rethinking programming language grammar towards efficient code generation}. In \bibinfo{booktitle}{\emph{Proceedings of the 33rd ACM SIGSOFT International Symposium on Software Testing and Analysis}}. \bibinfo{pages}{1124--1136}.
\newblock


\bibitem[\protect\citeauthoryear{Sun, Zhu, Xiong, Sun, Mou, and Zhang}{Sun et~al\mbox{.}}{2020}]%
        {sun2020treegen}
\bibfield{author}{\bibinfo{person}{Zeyu Sun}, \bibinfo{person}{Qihao Zhu}, \bibinfo{person}{Yingfei Xiong}, \bibinfo{person}{Yican Sun}, \bibinfo{person}{Lili Mou}, {and} \bibinfo{person}{Lu Zhang}.} \bibinfo{year}{2020}\natexlab{}.
\newblock \showarticletitle{Treegen: A tree-based transformer architecture for code generation}. In \bibinfo{booktitle}{\emph{Proceedings of the AAAI conference on artificial intelligence}}, Vol.~\bibinfo{volume}{34}. \bibinfo{pages}{8984--8991}.
\newblock


\bibitem[\protect\citeauthoryear{Vaswani, Shazeer, Parmar, Uszkoreit, Jones, Gomez, Kaiser, and Polosukhin}{Vaswani et~al\mbox{.}}{2017}]%
        {DBLP:conf/nips/VaswaniSPUJGKP17}
\bibfield{author}{\bibinfo{person}{Ashish Vaswani}, \bibinfo{person}{Noam Shazeer}, \bibinfo{person}{Niki Parmar}, \bibinfo{person}{Jakob Uszkoreit}, \bibinfo{person}{Llion Jones}, \bibinfo{person}{Aidan~N. Gomez}, \bibinfo{person}{Lukasz Kaiser}, {and} \bibinfo{person}{Illia Polosukhin}.} \bibinfo{year}{2017}\natexlab{}.
\newblock \showarticletitle{Attention is All you Need}. In \bibinfo{booktitle}{\emph{Advances in Neural Information Processing Systems 30: Annual Conference on Neural Information Processing Systems 2017, December 4-9, 2017, Long Beach, CA, {USA}}}, \bibfield{editor}{\bibinfo{person}{Isabelle Guyon}, \bibinfo{person}{Ulrike von Luxburg}, \bibinfo{person}{Samy Bengio}, \bibinfo{person}{Hanna~M. Wallach}, \bibinfo{person}{Rob Fergus}, \bibinfo{person}{S.~V.~N. Vishwanathan}, {and} \bibinfo{person}{Roman Garnett}} (Eds.). \bibinfo{pages}{5998--6008}.
\newblock
\urldef\tempurl%
\url{https://proceedings.neurips.cc/paper/2017/hash/3f5ee243547dee91fbd053c1c4a845aa-Abstract.html}
\showURL{%
\tempurl}


\bibitem[\protect\citeauthoryear{Wang, Li, Qian, Yang, Wang, Shang, Kumar, Tan, Ray, Bhatia, et~al\mbox{.}}{Wang et~al\mbox{.}}{2023b}]%
        {wang2023recode}
\bibfield{author}{\bibinfo{person}{Shiqi Wang}, \bibinfo{person}{Zheng Li}, \bibinfo{person}{Haifeng Qian}, \bibinfo{person}{Chenghao Yang}, \bibinfo{person}{Zijian Wang}, \bibinfo{person}{Mingyue Shang}, \bibinfo{person}{Varun Kumar}, \bibinfo{person}{Samson Tan}, \bibinfo{person}{Baishakhi Ray}, \bibinfo{person}{Parminder Bhatia}, {et~al\mbox{.}}} \bibinfo{year}{2023}\natexlab{b}.
\newblock \showarticletitle{ReCode: Robustness Evaluation of Code Generation Models}. In \bibinfo{booktitle}{\emph{The 61st Annual Meeting Of The Association For Computational Linguistics}}.
\newblock


\bibitem[\protect\citeauthoryear{Wang, Le, Gotmare, Bui, Li, and Hoi}{Wang et~al\mbox{.}}{2023a}]%
        {wang2023codet5+}
\bibfield{author}{\bibinfo{person}{Yue Wang}, \bibinfo{person}{Hung Le}, \bibinfo{person}{Akhilesh~Deepak Gotmare}, \bibinfo{person}{Nghi~DQ Bui}, \bibinfo{person}{Junnan Li}, {and} \bibinfo{person}{Steven~CH Hoi}.} \bibinfo{year}{2023}\natexlab{a}.
\newblock \showarticletitle{Codet5+: Open code large language models for code understanding and generation}.
\newblock \bibinfo{journal}{\emph{arXiv preprint arXiv:2305.07922}} (\bibinfo{year}{2023}).
\newblock


\bibitem[\protect\citeauthoryear{Wang, Wang, Joty, and Hoi}{Wang et~al\mbox{.}}{2021}]%
        {wang2021codet5}
\bibfield{author}{\bibinfo{person}{Yue Wang}, \bibinfo{person}{Weishi Wang}, \bibinfo{person}{Shafiq Joty}, {and} \bibinfo{person}{Steven~CH Hoi}.} \bibinfo{year}{2021}\natexlab{}.
\newblock \showarticletitle{Codet5: Identifier-aware unified pre-trained encoder-decoder models for code understanding and generation}.
\newblock \bibinfo{journal}{\emph{arXiv preprint arXiv:2109.00859}} (\bibinfo{year}{2021}).
\newblock


\bibitem[\protect\citeauthoryear{Wei, Bosma, Zhao, Guu, Yu, Lester, Du, Dai, and Le}{Wei et~al\mbox{.}}{2021}]%
        {wei2021finetuned}
\bibfield{author}{\bibinfo{person}{Jason Wei}, \bibinfo{person}{Maarten Bosma}, \bibinfo{person}{Vincent~Y Zhao}, \bibinfo{person}{Kelvin Guu}, \bibinfo{person}{Adams~Wei Yu}, \bibinfo{person}{Brian Lester}, \bibinfo{person}{Nan Du}, \bibinfo{person}{Andrew~M Dai}, {and} \bibinfo{person}{Quoc~V Le}.} \bibinfo{year}{2021}\natexlab{}.
\newblock \showarticletitle{Finetuned language models are zero-shot learners}.
\newblock \bibinfo{journal}{\emph{arXiv preprint arXiv:2109.01652}} (\bibinfo{year}{2021}).
\newblock


\bibitem[\protect\citeauthoryear{Wingate, Shoeybi, and Sorensen}{Wingate et~al\mbox{.}}{2022}]%
        {wingate2022prompt}
\bibfield{author}{\bibinfo{person}{David Wingate}, \bibinfo{person}{Mohammad Shoeybi}, {and} \bibinfo{person}{Taylor Sorensen}.} \bibinfo{year}{2022}\natexlab{}.
\newblock \showarticletitle{Prompt compression and contrastive conditioning for controllability and toxicity reduction in language models}.
\newblock \bibinfo{journal}{\emph{arXiv preprint arXiv:2210.03162}} (\bibinfo{year}{2022}).
\newblock


\bibitem[\protect\citeauthoryear{Xia, Deng, and Zhang}{Xia et~al\mbox{.}}{2024}]%
        {xia2024top}
\bibfield{author}{\bibinfo{person}{Chunqiu~Steven Xia}, \bibinfo{person}{Yinlin Deng}, {and} \bibinfo{person}{Lingming Zhang}.} \bibinfo{year}{2024}\natexlab{}.
\newblock \showarticletitle{Top Leaderboard Ranking= Top Coding Proficiency, Always? EvoEval: Evolving Coding Benchmarks via LLM}.
\newblock \bibinfo{journal}{\emph{arXiv preprint arXiv:2403.19114}} (\bibinfo{year}{2024}).
\newblock


\bibitem[\protect\citeauthoryear{Yang, Zhou, Yang, Yue, Chen, and Chen}{Yang et~al\mbox{.}}{2024b}]%
        {yang2024important}
\bibfield{author}{\bibinfo{person}{Guang Yang}, \bibinfo{person}{Yu Zhou}, \bibinfo{person}{Wenhua Yang}, \bibinfo{person}{Tao Yue}, \bibinfo{person}{Xiang Chen}, {and} \bibinfo{person}{Taolue Chen}.} \bibinfo{year}{2024}\natexlab{b}.
\newblock \showarticletitle{How important are good method names in neural code generation? a model robustness perspective}.
\newblock \bibinfo{journal}{\emph{ACM Transactions on Software Engineering and Methodology}} \bibinfo{volume}{33}, \bibinfo{number}{3} (\bibinfo{year}{2024}), \bibinfo{pages}{1--35}.
\newblock


\bibitem[\protect\citeauthoryear{Yang, Sun, Yue, Devanbu, and Lo}{Yang et~al\mbox{.}}{2024a}]%
        {yang2024robustness}
\bibfield{author}{\bibinfo{person}{Zhou Yang}, \bibinfo{person}{Zhensu Sun}, \bibinfo{person}{Terry~Zhuo Yue}, \bibinfo{person}{Premkumar Devanbu}, {and} \bibinfo{person}{David Lo}.} \bibinfo{year}{2024}\natexlab{a}.
\newblock \showarticletitle{Robustness, security, privacy, explainability, efficiency, and usability of large language models for code}.
\newblock \bibinfo{journal}{\emph{arXiv preprint arXiv:2403.07506}} (\bibinfo{year}{2024}).
\newblock


\bibitem[\protect\citeauthoryear{Yin and Neubig}{Yin and Neubig}{2017}]%
        {yin2017syntactic}
\bibfield{author}{\bibinfo{person}{Pengcheng Yin} {and} \bibinfo{person}{Graham Neubig}.} \bibinfo{year}{2017}\natexlab{}.
\newblock \showarticletitle{A syntactic neural model for general-purpose code generation}.
\newblock \bibinfo{journal}{\emph{arXiv preprint arXiv:1704.01696}} (\bibinfo{year}{2017}).
\newblock


\bibitem[\protect\citeauthoryear{Zheng, Xia, Zou, Dong, Wang, Xue, Wang, Shen, Wang, Li, Su, Yang, and Tang}{Zheng et~al\mbox{.}}{2023}]%
        {zheng2023codegeex}
\bibfield{author}{\bibinfo{person}{Qinkai Zheng}, \bibinfo{person}{Xiao Xia}, \bibinfo{person}{Xu Zou}, \bibinfo{person}{Yuxiao Dong}, \bibinfo{person}{Shan Wang}, \bibinfo{person}{Yufei Xue}, \bibinfo{person}{Zihan Wang}, \bibinfo{person}{Lei Shen}, \bibinfo{person}{Andi Wang}, \bibinfo{person}{Yang Li}, \bibinfo{person}{Teng Su}, \bibinfo{person}{Zhilin Yang}, {and} \bibinfo{person}{Jie Tang}.} \bibinfo{year}{2023}\natexlab{}.
\newblock \showarticletitle{CodeGeeX: A Pre-Trained Model for Code Generation with Multilingual Benchmarking on HumanEval-X}. In \bibinfo{booktitle}{\emph{Proceedings of the 29th ACM SIGKDD Conference on Knowledge Discovery and Data Mining}}. \bibinfo{pages}{5673--5684}.
\newblock


\bibitem[\protect\citeauthoryear{Zhuo, Vu, Chim, Hu, Yu, Widyasari, Yusuf, Zhan, He, Paul, et~al\mbox{.}}{Zhuo et~al\mbox{.}}{2024}]%
        {zhuo2024bigcodebench}
\bibfield{author}{\bibinfo{person}{Terry~Yue Zhuo}, \bibinfo{person}{Minh~Chien Vu}, \bibinfo{person}{Jenny Chim}, \bibinfo{person}{Han Hu}, \bibinfo{person}{Wenhao Yu}, \bibinfo{person}{Ratnadira Widyasari}, \bibinfo{person}{Imam Nur~Bani Yusuf}, \bibinfo{person}{Haolan Zhan}, \bibinfo{person}{Junda He}, \bibinfo{person}{Indraneil Paul}, {et~al\mbox{.}}} \bibinfo{year}{2024}\natexlab{}.
\newblock \showarticletitle{Bigcodebench: Benchmarking code generation with diverse function calls and complex instructions}.
\newblock \bibinfo{journal}{\emph{arXiv preprint arXiv:2406.15877}} (\bibinfo{year}{2024}).
\newblock


\end{thebibliography}
